\documentclass[%
 preprint,
 amsmath,amssymb,
 aps, physrev,
]{revtex4-2}
\usepackage{graphicx}
\usepackage{svg}
\usepackage[T1]{fontenc}
\usepackage{dcolumn}
\usepackage{bm}
\usepackage{ulem}
\usepackage{soul}
\usepackage{verbatim}
\usepackage{multirow}
\usepackage{xcolor}
\usepackage[version=4]{mhchem}

\DeclareUnicodeCharacter{2500}{}
\begin{document}
\preprint{APS/123-QED}

\title{\textbf{
Exact Mirror-Parity Selection Rule for Electron–Phonon Coupling in Two-Dimensional Crystals} 
}%

\author{Nicolás F. Barrera}
\affiliation{Laboratoire de Chimie Théorique, Sorbonne Université and CNRS, 4 Pl Jussieu, F-75005,
Paris, France.}
\author{Javiera Cabezas-Escares}
\affiliation{
Departamento de Física, Facultad de Ciencias, Universidad de Chile and Center for Nanoscience and Nanotechnology (CEDENNA) Santiago, Chile.
}
\author{Francisco Munoz}
\affiliation{
Departamento de Física, Facultad de Ciencias, Universidad de Chile and Center for Nanoscience and Nanotechnology (CEDENNA) Santiago, Chile.
}

\author{Carlos Cardenas}
\email{Contact author: cardena@uchile.cl}
\affiliation{
Departamento de Física, Facultad de Ciencias, Universidad de Chile.
}
\affiliation{
Center for Nanoscience and Nanotechnology (CEDENNA) Santiago, Chile.
}
\affiliation{
Universidad Nacional de Colombia - Sede Medell\'in, Facultad de Minas,     Departamento de Materiales y Minerales. Carrera 80 \# 65-223,     N\'ucleo Robledo, Medell\'in, Colombia
}

\author{Julia Contreras-García}%
\homepage{https://www.lct.jussieu.fr/pagesperso/contrera/}
\email{Contact author: julia.contreras\_garcia@sorbonne-universite.fr}
\affiliation{Laboratoire de Chimie Théorique, Sorbonne Université and CNRS, 4 Pl Jussieu, F-75005,
Paris, France.
}



\date{\today}

\begin{abstract}
The electron--phonon (e-ph) interaction underlies conventional
superconductivity and plays a central role in transport and other
many-body phenomena in solids. Here, we derive a symmetry selection
rule for e-ph scattering in two-dimensional crystals possessing a
horizontal mirror plane, $\sigma_h$. For electronic states of mirror
parity $s_n(\mathbf{k})$ and $s_m(\mathbf{k}+\mathbf{q})$, and a
phonon mode of parity $\chi_\nu(\mathbf{q})$, we show that the e-ph
matrix element $g_{mn,\nu}(\mathbf{k},\mathbf{q})$ vanishes identically
unless $s_n(\mathbf{k})\,\chi_\nu(\mathbf{q})\,
s_m(\mathbf{k}+\mathbf{q})=+1$. The rule follows from the covariance of
the phonon-induced deformation potential under $\sigma_h$, constitutes
the $C_s$ case of the general group-theoretical selection rule for
matrix elements, and partitions the e-ph interaction throughout the
Brillouin zone into symmetry-allowed and symmetry-forbidden scattering
channels. Consequently, mirror-even phonons couple electronic states of
the same mirror parity, either even--even or odd--odd, whereas
mirror-odd phonons couple only states of opposite mirror parity, for
both intra- and interband processes.
We illustrate the selection rule through first-principles calculations
for the two-dimensional superconductor \ce{Sc2BCH4}, where mirror
parity is well defined throughout the entire Brillouin zone and
symmetry-forbidden matrix elements vanish up to numerical precision.
We then extend the analysis to bulk \ce{MgB2}, where the rule applies
exactly only on mirror-invariant planes of the three-dimensional Brillouin zone. Moving away from these planes removes the exact mirror constraint and activates scattering channels that are symmetry forbidden when the mirror eigenvalues are well defined. Our results establish horizontal mirror symmetry as an exact and practical filter of e-ph scattering processes. 
\end{abstract}

\maketitle


\section{Introduction}
\label{sec:intro}

Interactions between fermions and bosonic excitations are a central
ingredient of many-body physics, where they provide scattering channels,
renormalize quasiparticle properties, and can mediate effective
interactions between fermions. In solids, the electron--phonon (e-ph)
interaction is a paradigmatic example, underlying phenomena ranging
from electrical and thermal transport to charge-density-wave
instabilities and conventional superconductivity
\cite{RevModPhys.89.015003}. In the latter, Bardeen--Cooper--Schrieffer (BCS) theory established
that the key properties of conventional superconductivity can be
understood in terms of Cooper pairing driven by an effective attractive
interaction between electrons mediated by the exchange of phonons
\cite{frohlich1950theory,bardeen1955electron,bardeen1957theory}. Migdal--Eliashberg (ME) theory subsequently
provided a quantitative framework for describing this interaction in
real materials, including its frequency dependence and the
renormalization of the electronic quasiparticles
\cite{migdal1958interaction,eliashberg1960interactions}. As a result, an accurate evaluation
of the e-ph interaction has become a cornerstone of first-principles
predictions of the superconducting critical temperature in
conventional superconductors.

The development of density-functional perturbation theory and
efficient Brillouin-zone interpolation techniques has made such
calculations practical for increasingly complex materials
\cite{RevModPhys.73.515,giustino2007electron,RevModPhys.89.015003}. Implementations in
first-principles packages such as Quantum ESPRESSO\cite{10.1063/5.0005082}, EPW\cite{lee2023electron}, Abinit\cite{10.1063/5.0288278}, and CASTEP\cite{clark2005first}
now provide direct access to phonons, e-ph matrix elements, and
Migdal--Eliashberg superconducting properties. Combined with modern
crystal-structure prediction methods, including evolutionary,
random-search, and swarm-optimization algorithms
\cite{glass2006uspex,hajinazar2025xtalopt,PhysRevLett.97.045504}, these developments have changed the search
for conventional superconductors from the analysis of known compounds
to the prediction of new ones. High-pressure hydrides provide the
clearest examples. Structure searches and first-principles e-ph
calculations anticipated high-$T_c$ superconductivity in H$_3$S and
LaH$_{10}$, followed by experimental observations near 203 and
250 K, respectively
\cite{duan2014pressure,drozdov2015conventional,doi:10.1073/pnas.1704505114,drozdov2019superconductivity}.

The record transition temperatures of hydrogen-rich compounds,
however, are obtained at megabar pressures, which severely limits
their accessibility and motivates the search for materials that retain
strong phonon-mediated pairing under ambient conditions
\cite{gao2025maximum,sanna2026search,pires2026machine}. Two-dimensional materials offer a distinct route to overcome these
limitations. Unlike bulk hydrides, whose stability is often tied to
extreme pressure, atomically thin phases can be isolated or kinetically
stabilized through mechanical and in situ exfoliation, substrate-assisted
growth, and molecular-beam epitaxy, providing access to structures that
do not necessarily correspond to the thermodynamic ground state of the bulk
\cite{novoselov2004electric,huang2020universal,seredynski2021molecular,grubivsic2023situ}.
Once stabilized, their exposed surfaces provide an additional degree of
control through chemical functionalization\cite{li2021synthesis}. In this context, hydrogenation has emerged as a promising strategy for
enhancing conventional superconductivity in two-dimensional materials.
The seminal work of Bekaert \textit{et al.}\cite{bekaert2019hydrogen}
showed that hydrogenating  a monolayer of \ce{MgB2} can raise its  $T_c$ to
67 K at ambient pressure and to nearly 100 K under biaxial strain,
demonstrating that hydrogen functionalization can strongly enhance the
e-ph interaction in a low-dimensional host. This idea has since been
extended to a wide range of two-dimensional borides and related
compounds, where hydrogenation, strain, and chemical composition have
been used to engineer electronic states at the Fermi level and the
phonon spectrum, leading to predicted values $T_c$ approaching or
exceeding 100 K at ambient pressure
\cite{sevik2022high,han2023theoretical,seeyangnok2025high,
xie2025hydrogenation,soskic2024enhanced}.

Beyond expanding the pool of candidate materials for
ambient-pressure superconductivity, reduced dimensionality also
provides a particularly simple setting in which crystal symmetry can
constrain the e-ph interaction\cite{PhysRevB.106.045301,milovsevic2019symmetry,nikolic2021electron,shu2024full}. 
In a two-dimensional crystal that has a horizontal mirror plane $\sigma_h$, the mirror operation leaves all crystal momenta in the plane unchanged. Electronic Bloch states and phonon modes can therefore be chosen as eigenstates of $\sigma_h$ throughout the Brillouin zone and classified according to their corresponding mirror eigenvalues. In this work,  we derive general selection rules for the  electron--phonon
scattering problem.

A particularly close precursor to the present work was given by
Alidoosti \textit{et al.}\cite{PhysRevB.106.045301}, who showed that
the transformation of the phonon-induced deformation potential under
$\sigma_h$ determines whether flexural phonons can couple linearly to
electronic states. In the long-wavelength limit and for a minimal
two-atom model, they demonstrated that when the atoms lie in the
mirror plane, flexural distortions generate a mirror-odd deformation
potential, and the corresponding diagonal e-ph matrix elements vanish. When atoms lie away from the mirror plane in pairs related by $\sigma_h$, their out-of-plane displacements can combine with different mirror symmetries. The acoustic flexural mode is odd under $\sigma_h$, whereas an optical flexural mode can be even. The latter can therefore couple to electronic states through the e-ph  interaction. Their analysis thus established the essential
role of the mirror character of the deformation potential, rather than
the out-of-plane character of the atomic displacement itself.

In this work, we extend this idea to the general electron--phonon
scattering problem. Rather than restricting the analysis to diagonal
matrix elements or particular phonon modes, we derive a selection rule
for arbitrary initial and final Bloch states and any phonon mode at any
wave vector in a two-dimensional Brillouin zone. We show that the  electron--phonon matrix,
$g_{mn,\nu}(\mathbf{k},\mathbf{q})$, vanishes identically unless
\begin{equation}
s_n(\mathbf{k})\,\chi_\nu(\mathbf{q})\,
s_m(\mathbf{k}+\mathbf{q})=+1.
\end{equation}
where $s_n(\mathbf{k})$, $s_m(\mathbf{k}+\mathbf{q})$, and $\chi_\nu(\mathbf{q})$
are the $\hat{\sigma}_h$ eigenvalues ($\pm1$), i.e., the mirror parities,
of Kohn--Sham states and of the phonon-induced
perturbation $\Delta_{\mathbf{q}\nu}v^{\rm KS}$, respectively. Thus, mirror-even phonons couple electronic states of the same mirror
parity, whereas mirror-odd phonons couple states of opposite parity,
for both intra- and interband scattering processes. The rule follows
from the covariance of the phonon-induced deformation potential under
$\sigma_{h}$ and can equivalently be viewed as the $C_{s}$ case of the
general group-theoretical selection rule for matrix elements~\cite{bradley1972mathematical,dresselhaus2008group}, as
discussed in Sec.~III. It partitions the electron--phonon interaction
into symmetry-allowed and symmetry-forbidden scattering channels
throughout the two-dimensional Brillouin zone.

We demonstrate the consequences of this rule through first-principles density-functional perturbation theory
(DFPT) calculations for two archetypical superconductors. We first
consider the two-dimensional \ce{Sc2BCH4} monolayer, for which
$\sigma_h$ belongs to the little group of every  wave vector in the Brillouin zone
. This system therefore provides a direct realization of the selection rule throughout the entire Brillouin
zone. We then turn to bulk \ce{MgB2} to examine how the same symmetry argument operates in a layered three-dimensional crystal. In this case, mirror parity is
well defined only on mirror-invariant planes of the Brillouin zone,
where the selection rule remains exact. By continuously moving the
electronic wave vector away from those planes, we further show how
scattering channels that are strictly forbidden by symmetry become
active once the electronic states no longer possess well-defined
mirror eigenvalues. Together, these examples provide an explicit
first-principles demonstration of the rule across the entire
two-dimensional Brillouin zone, establish horizontal mirror symmetry as
an exact and practical filter of electron--phonon scattering channels,
and clarify the dimensional conditions under which this constraint
applies.

The remainder of this article is organized as follows. Section~II
introduces the theoretical framework for the electron--phonon
interaction and the symmetry properties required for the analysis.
Section~III derives the mirror-parity selection rule and discusses the
corresponding symmetry-allowed and symmetry-forbidden scattering
channels. Section~IV presents the numerical results for \ce{Sc2BCH4}
and \ce{MgB2}. Section~V describes the computational details, and
Sec.~VI summarizes our main conclusions.

\section{Theoretical background}
\label{sec:theory_background}
To establish the symmetry constraints governing the electron--phonon interaction, we first introduce the standard microscopic framework and fix the notation used throughout this work. Within the harmonic approximation for the lattice and retaining the electron--phonon interaction to first order in the ionic displacements, the coupled electron--phonon system can be described by the following Hamiltonian:
\begin{equation}
\label{eq:H_sq}
\hat{H} = \sum_{n\mathbf{k}}
\varepsilon_{n\mathbf{k}}
\hat{c}_{n\mathbf{k}}^{\dagger}
\hat{c}_{n\mathbf{k}}
+
\sum_{\mathbf{q}\nu}
\hbar\omega_{\mathbf{q}\nu}
\left(
\hat{a}^{\dagger}_{\mathbf{q}\nu}
\hat{a}_{\mathbf{q}\nu}
+\frac{1}{2}
\right)+
\frac{1}{\sqrt{N_p}}
\sum_{\substack{\mathbf{k},\mathbf{q}\\mn\nu}}
g_{mn\nu}(\mathbf{k},\mathbf{q})\,
\hat{c}^{\dagger}_{m,\mathbf{k}+\mathbf{q}}
\hat{c}_{n\mathbf{k}}
\left(
\hat{a}_{\mathbf{q}\nu}
+
\hat{a}^{\dagger}_{-\mathbf{q}\nu}
\right).
\end{equation}
Here, $\varepsilon_{n\mathbf{k}}$ is the eigenvalue of an electronic state in band $n$ with crystal momentum $\mathbf{k}$, and $\omega_{\mathbf{q}\nu}$ is the frequency of a phonon with wave vector $\mathbf{q}$ and branch index $\nu$.
The operators $\hat{c}_{n\mathbf{k}}^{(\dagger)}$ and $\hat{a}_{\mathbf{q}\nu}^{(\dagger)}$ are the corresponding fermionic and bosonic annihilation (creation) operators, respectively, and $N_p$ is the number of primitive cells in the Born--von K\'arm\'an supercell. The e-ph matrix element $g_{mn\nu}(\mathbf{k},\mathbf{q})$ describes the scattering of an electron from $(n,\mathbf{k})$ to $(m,\mathbf{k}+\mathbf{q})$ through the absorption or emission of a phonon $(\mathbf{q},\nu)$.

We next outline how the three contributions in Eq.~\eqref{eq:H_sq} arise within a first-principles description based on density-functional theory (DFT) and density-functional perturbation theory (DFPT). We focus on the quantities and transformation properties required for the symmetry analysis developed in
Sec.~\ref{sec:theory}.

\subsection{Electronic structure}
\label{subsec:elec_struc}
In DFT, the interacting many-electron problem is mapped onto an
auxiliary Kohn--Sham (KS) system of noninteracting electrons    that
reproduces the ground-state electron density. The KS states $\psi_{n\mathbf{k}}(\mathbf r)$ satisfy
\begin{equation}
\label{eq:KS}
\hat{H}_{\rm KS}\psi_{n\mathbf{k}}(\mathbf r)
=
\varepsilon_{n\mathbf{k}}\psi_{n\mathbf{k}}(\mathbf r),
\qquad
\hat{H}_{\rm KS}
=
-\frac{\hbar^2}{2m}\nabla^2+V_{\rm KS},
\end{equation}
where $V_{\rm KS}$ denotes the effective self-consistent KS potential
operator, including the ionic, Hartree, and exchange--correlation
contributions and, when present, nonlocal pseudopotential or
Hubbard-$U$ terms. For a periodic crystal,  the KS states be chosen as Bloch states,
\begin{equation}
\label{eq:bloch}
\psi_{n\mathbf{k}}(\mathbf r)
=
\dfrac{1}{\sqrt{N_p}}
u_{n\mathbf{k}}(\mathbf r)
e^{i\mathbf{k}\cdot\mathbf r},
\qquad
u_{n\mathbf{k}}(\mathbf r+\mathbf R)
=
u_{n\mathbf{k}}(\mathbf r),
\end{equation}
where $\mathbf R$ is a lattice vector and
$u_{n\mathbf{k}}(\mathbf r)$ is lattice-periodic and normalized over
the primitive cell. Following the standard second-quantization prescription, the effective one-body KS Hamiltonian can be written as
\begin{equation}
\hat{H}_{e} = \sum_{n\mathbf{k},n'\mathbf{k'}}\langle\psi_{n\mathbf{k}}|\hat{H}_{KS}|\psi_{n'\mathbf{k'}}\rangle\hat{c}_{n\mathbf{k}}^{\dagger}\hat{c}_{n'\mathbf{k'}}  =  \sum_{n\mathbf{k}}\varepsilon_{n\mathbf{k}}\hat{c}_{n\mathbf{k}}^{\dagger}\hat{c}_{n\mathbf{k}},
\end{equation}
where the orthonormality of the KS eigenstates has been used. This
recovers the first term of Eq.~\eqref{eq:H_sq}.
\subsection{Lattice vibrations}
\label{subsec:lattice_vib}

Within the harmonic approximation, the lattice displacements are
resolved into independent normal modes labeled by the phonon wave
vector $\mathbf q$ and branch index $\nu$. The displacement of atom
$\kappa$ along the 
Cartesian direction $\alpha$ in cell $\mathbf R_p$,
quantized in terms of phonon creation and annihilation operators, is
\begin{equation}
\label{eq:displacement}
\Delta\tau_{\kappa\alpha p}
=
\left(
\frac{\hbar}{2N_pM_\kappa}
\right)^{1/2}
\sum_{\mathbf q\nu}
\frac{e_{\kappa\alpha,\nu}(\mathbf q)}{\sqrt{\omega_{\mathbf q\nu}}}\,
e^{i\mathbf q\cdot\mathbf R_p}
\left(
\hat{a}_{\mathbf q\nu}
+
\hat{a}_{-\mathbf q\nu}^{\dagger}
\right),
\end{equation}
where $M_\kappa$ is the mass of atom $\kappa$ and
$e_{\kappa\alpha,\nu}(\mathbf q)$ is the
polarization vector of phonon branch $\nu$, i.e., an eigenvector of
the mass-weighted dynamical matrix.   The polarization vectors are
normalized according to  $\sum_{\kappa\alpha}|e_{\kappa\alpha,\nu}(\mathbf q)|^2=1$.  Equation~\eqref{eq:displacement} adopts the phase convention $e_{\kappa\alpha,\nu}(-\mathbf q)=e_{\kappa\alpha,\nu}^{*}(\mathbf q)$.

In this  normal-mode basis, the harmonic lattice Hamiltonian is diagonal,
\begin{equation}
\label{eq:Hph}
\hat{H}_{\rm ph}
=
\sum_{\mathbf q\nu}
\hbar\omega_{\mathbf q\nu}
\left(
\hat{a}_{\mathbf q\nu}^{\dagger}
\hat{a}_{\mathbf q\nu}
+\frac{1}{2}
\right),
\end{equation}
recovering the second term of Eq.~\eqref{eq:H_sq}. Within DFPT, the frequencies  and polarization vectors follow from the linear response of  the ground state to periodic atomic displacements
\cite{RevModPhys.73.515}.

\subsection{Electron--phonon coupling}
\label{subsec:e_ph_coupling}

The e-ph interaction originates from the dependence of the
effective KS potential on the ionic positions. For small
displacements $\Delta\tau_{\kappa\alpha p}$ around equilibrium
$\{\tau^0\}$, we expand $V_{\rm KS}$ to the first order,
\begin{equation}
\label{eq:VKS_expansion}
V_{\rm KS}(\{\tau_{\kappa p}\})
=
V_{\rm KS}(\{\tau^0_{\kappa p}\})
+
\sum_{\kappa\alpha p}
\left.\frac{\partial V_{\rm KS}}
{\partial \tau_{\kappa\alpha p}}\right|_{\{\tau^0\}}
\Delta\tau_{\kappa\alpha p}
+
\mathcal{O}(\Delta\tau^2).
\end{equation}
This expression can be rewritten into normal mode coordinates using Eq.\ref{eq:displacement}:
\begin{equation}
\label{eq:VKS_normal}
\Delta V^{\mathrm{KS}} =V_{\rm KS}(\{\tau_{\kappa p}\})
-
V_{\rm KS}(\{\tau^0_{\kappa p}\}) = \dfrac{1}{\sqrt{N_{p}}}\sum_{\mathbf{q}\nu}\Delta_{\mathbf{q}\nu}V^{\mathrm{KS}}(\hat{a}_{\mathbf{q},\nu}+\hat{a}^{\dagger}_{-\mathbf{q},\nu}),   
\end{equation}
where
\begin{equation}
\Delta_{\mathbf{q}\nu}V^{\mathrm{KS}} = \bigg(\dfrac{\hbar}{2\omega_{\mathbf{q\nu}}}\bigg)^{1/2}\sum_{\kappa\alpha} \dfrac{e_{\kappa\alpha,\nu}(\mathbf{q})}{\sqrt{M_{\kappa}}}\sum_{p}e^{i\mathbf{q}\cdot\mathbf{R}_{p}}\dfrac{\partial V^{\mathrm{KS}}(\mathbf{r})}{\partial\tau_{\kappa\alpha p}}\bigg|_{\tau^{0}}.
\label{eq:Delta_qv}
\end{equation}
Owing to the translational invariance of the crystal, the derivatives
of the KS potential transform covariantly under lattice translations,
\begin{equation}
\left.\frac{\partial V^{\mathrm{KS}}(\mathbf{r})}{\partial\tau_{\kappa\alpha p}}\right|_{\tau^0}
= 
\partial_{\kappa\alpha0}V^{\mathrm{KS}}(\mathbf{r}-\mathbf{R}_{p}),
\end{equation}
the sum over $p$ in $\Delta_{\mathbf{q}\nu}V^{\mathrm{KS}}$ takes the form of a Bloch sum, so that
\begin{equation}
\label{eq:bloch_factorization}
\Delta_{\mathbf{q}\nu}V^{\mathrm{KS}}(\mathbf{r}) = e^{i\mathbf{q}\cdot\mathbf{r}}\,\Delta_{\mathbf{q}\nu}v^{\mathrm{KS}}(\mathbf{r}),
\end{equation}
where $\Delta_{\mathbf{q}\nu}v^{\mathrm{KS}}(\mathbf{r})$ is periodic under the lattice,
$\Delta_{\mathbf{q}\nu}v^{\mathrm{KS}}(\mathbf{r}+\mathbf{R}_{p})=\Delta_{\mathbf{q}\nu}v^{\mathrm{KS}}(\mathbf{r})$.

The transition to second quantization yields to:
\begin{eqnarray}
\hat{H}_{e-ph} = \sum_{n\mathbf{k},n'\mathbf{k'}}\langle\psi_{n\mathbf{k}}|\Delta V^{\mathrm{KS}}|\psi_{n'\mathbf{k}'}\rangle\hat{c}^{\dagger}_{n\mathbf{k}}\hat{c}_{n'\mathbf{k}'}.
\end{eqnarray}
Using Eqs. \eqref{eq:bloch} and \eqref{eq:VKS_normal}-\eqref{eq:bloch_factorization}, the electron-phonon coupling Hamiltonian can be rewritten as:
\begin{equation}
\hat{H}_{e-ph} = \dfrac{1}{\sqrt{N_{p}}}\sum_{\substack{\mathbf{k},\mathbf{q}\\mn\nu}}g_{mn\nu}(\mathbf{k},\mathbf{q})\hat{c}^{\dagger}_{m,\mathbf{k}+\mathbf{q}}\hat{c}_{n,\mathbf{k}}(\hat{a}_{\mathbf{q},\nu}+\hat{a}^{\dagger}_{-\mathbf{q},\nu}),
\end{equation}
where the electron-phonon matrix element is given by
\begin{equation}
\label{eq:g_matrix_element}
g_{mn\nu}(\mathbf{k},\mathbf{q}) = \langle u_{m,\mathbf{k}+\mathbf{q}}|\Delta_{\mathbf{q},\nu}v^{\mathrm{KS}}|u_{n_\mathbf{k}}\rangle_{uc},
\end{equation}
where the subscript $uc$ denotes integration over the primitive unit cell. Equation~\eqref{eq:g_matrix_element} is the central quantity considered in this work. Its symmetry properties are controlled by the transformation of the initial and final electronic states and of the
phonon-induced perturbation $\Delta v_{\nu,\mathbf q}$. In the following, we show that horizontal mirror symmetry imposes an exact selection rule on this matrix element.

\section{Mirror-parity selection rule}
\label{sec:theory}
\subsection{Covariance of the Kohn-Sham potential}

We define $\sigma_{h}$ as the horizontal mirror plane of the two-dimensional crystal, $(x,y,z)\mapsto(x,y,-z)$, and $\hat{\sigma}_{h}$ as its unitary  representation in $\mathcal{H}_{1}$:  $(\hat{\sigma}_{h}\psi)(\mathbf{r})=\psi(\sigma_{h}^{-1}\mathbf{r})$. This operation permutes the atoms of the primitive cell,  $\kappa\mapsto\sigma_h(\kappa)$, and assigns an intrinsic parity $\eta_{\alpha}$ to each Cartesian direction ($\eta_{x}=\eta_{y}=+1$, $\eta_{z}=-1$).

The Kohn--Sham potential is the functional derivative of the energy terms that 
depend explicitly  on the density $n$ (Hartree, exchange--correlation, and external) and of the 
non-local terms. Since these terms are invariant under the joint transformation 
$(\mathbf r,n,\{\tau\})\mapsto(\sigma_h\mathbf r,\,n\circ\sigma_h^{-1},\,
\{\sigma_h\tau\})$, the potential must transform consistently:
\begin{equation}
\label{eq:cov_KS}
    \hat\sigma_h\,\hat V^{\mathrm{KS}}[\{\tau\}]\,\hat\sigma_h^{-1} 
    = \hat V^{\mathrm{KS}}[\{\sigma_h\tau\}],
\end{equation}
for any atomic configuration $\{\tau\}$, with 
$\{\sigma_h\tau\}_{\kappa\alpha}=\eta_\alpha\,\tau_{\sigma_h(\kappa),\alpha}$.

\subsection{Covariance of the deformation potential}
Starting with the equilibrium configuration  $\{\tau^{0}\}$, with a small displacement $\varepsilon$ on the $\kappa$ atom,  $\alpha$ direction and $p$ cell,
\begin{equation}
 \tau_{\kappa'\beta p'} = \tau^{0}_{\kappa'\beta p'} + \varepsilon\delta_{\kappa'\kappa}\delta_{\beta\alpha}\delta_{p'p}.
\end{equation}
Operating with $\sigma_h$ and using that the equilibrium position is invariant under the specular horizontal plane, the transformation of the deformed atomic position reads,
\begin{equation}
\begin{split}
    \{\sigma_{h}\tau\}_{\kappa'\beta p'}  =
\tau^{0}_{\kappa'\beta p'} + \varepsilon\eta_{\alpha}\delta_{\kappa',\sigma_{h}(\kappa)}\delta_{\beta\alpha}\delta_{p'p} .
\end{split}
\end{equation}

And the transformation of the potential (Eq.~\eqref{eq:cov_KS})  can be rewritten as:
\begin{equation}
\hat{\sigma}_{h}\hat{V}^{\mathrm{KS}}[\tau^{0}_{\kappa'\beta p'} + \varepsilon\delta_{\kappa'\kappa}\delta_{\beta\alpha}\delta_{p'p}]\hat{\sigma}^{-1}_{h} = \hat{V}^{\mathrm{KS}}[\tau^{0}_{\kappa'\beta p'} + \varepsilon\eta_{\alpha}\delta_{\kappa',\sigma_{h}(\kappa)}\delta_{\beta\alpha}\delta_{p'p} ].
\end{equation}
To simplify the notation, the equilibrium position $\tau^0_{\kappa'\beta p'}$ will be called $\tau^0$.  Now differentiate both sides of this equation with respect to $\varepsilon$ and evaluating at $\varepsilon=0$,
\begin{equation}
\dfrac{d}{d\varepsilon}\Big[\hat{\sigma}_{h}\hat{V}^{\mathrm{KS}}[\tau^{0}+\varepsilon\delta_{\kappa'\kappa}\delta_{\beta\alpha}\delta_{p'p}]\hat{\sigma}^{-1}_{h}\Big]_{\varepsilon = 0} = \dfrac{d}{d\varepsilon}\Big[\hat{V}^{\mathrm{KS}}[\tau^{0} + \varepsilon\eta_{\alpha}\delta_{\kappa',\sigma_{h}(\kappa)}\delta_{\beta\alpha}\delta_{p'p}]\Big]_{\varepsilon = 0}.
\label{eq:dV/dq}
\end{equation}
Using the chain rule, the term $\frac{d\tau_{\kappa'\beta p'}}{d\varepsilon}$gives rise to Kronecker deltas, which collapse the sums,
\begin{equation}
\begin{split}
\hat{\sigma}_{h}\Big(\sum_{\kappa', \beta, p'} \frac{\partial \hat{V}^{\text{KS}}}{\partial \tau_{\kappa'\beta p'}} \cdot \delta_{\kappa'\kappa}\delta_{\beta\alpha}\delta_{p'p}\Big)\hat{\sigma}^{-1}_{h} &=\sum_{\kappa', \beta, p'} \frac{\partial \hat{V}^{\text{KS}}}{\partial \tau_{\kappa'\beta p'}} \cdot  \eta_{\alpha} \delta_{\kappa',\sigma_{h}(\kappa)}\delta_{\beta\alpha}\delta_{p'p},\\
\hat{\sigma}_{h}\Big(\partial_{\kappa\alpha p}\hat{V}^{\mathrm{KS}}\Big)\hat{\sigma}^{-1}_{h}  &=\eta_{\alpha} \partial_{{\sigma_{h}(\kappa)\alpha p}}\hat{V}^{\mathrm{KS}}.
\end{split}
\end{equation}

The last equation describes the effect of moving a specific atom $\kappa$ in a determinate cell $p$. To turn the effect on the perturbations, the Bloch phase for the wave vector $\mathbf{q}$ and the unit cell vector $\mathbf{R}_p(x,y) = \mathbf{R}_p$ are introduced:
\begin{equation}
e^{i\mathbf{q}\cdot\mathbf{R}_{p}}\hat{\sigma}_{h}\partial_{\kappa\alpha p}\hat{V}^{\mathrm{KS}}\hat{\sigma}_{h}^{-1} = e^{i\mathbf{q}\cdot\mathbf{R}_{p}}\eta_{\alpha}\partial_{\sigma_{h}(\kappa)\alpha p}\hat{V}^{\mathrm{KS}}. 
\end{equation}
Summing over the lattice cells, and considering that $e^{i\mathbf{q}\cdot\mathbf{R}_{p}}$ is invariant under $\sigma_h$ since $\mathbf{R}_{p}$ is purely two-dimensional:
\begin{equation}
\hat{\sigma}_{h}\bigg(\sum_{p}e^{i\mathbf{q}\cdot\mathbf{R}_{p}}\partial_{\kappa\alpha p}\hat{V}^{\mathrm{KS}}\bigg)\hat{\sigma}_{h}^{-1} = \eta_{\alpha}\sum_{p}e^{i\mathbf{q}\cdot\mathbf{R}_{p}}\partial_{\sigma_{h}(\kappa)\alpha p}\hat{V}^{\mathrm{KS}}. \label{eq:sum_p} 
\end{equation}
To recover the expression in terms of the normal mode coordinates for $\hat{V}^{\mathrm{KS}}$ in Eq.~\eqref{eq:Delta_qv}, we defined: 
\begin{equation}
\Delta_{\mathbf{q}}\partial_{\kappa\alpha}\hat{V}^{\mathrm{KS}} = \sum_{p}e^{i\mathbf{q}\cdot{\mathbf{R}_{p}}}\partial_{\kappa\alpha p}\hat{V}^{\mathrm{KS}}.
\end{equation}
Multiplying both sides of Eq.~\eqref{eq:sum_p} by
$(\hbar/2\omega_{\mathbf{q}\nu})^{1/2}e_{\kappa\alpha,\nu}(\mathbf{q})M_\kappa^{-1/2}$
and summing over $\kappa\alpha$ gives
\begin{equation}
\label{eq:dV_both_sides}
\bigg(\dfrac{\hbar}{2\omega_{\mathbf{q\nu}}}\bigg)^{1/2}\sum_{\kappa\alpha}\dfrac{e_{\kappa\alpha,\nu}(\mathbf{q})}{\sqrt{M_{\kappa}}}\hat{\sigma}_{h}\Delta_{\mathbf{q}}\partial_{\kappa\alpha}\hat{V}^{\mathrm{KS}}\hat{\sigma}_{h}^{-1} = \bigg(\dfrac{\hbar}{2\omega_{\mathbf{q\nu}}}\bigg)^{1/2}\sum_{\kappa\alpha}\dfrac{e_{\kappa\alpha,\nu}(\mathbf{q})}{\sqrt{M_{\kappa}}}\eta_{\alpha}\Delta_{\mathbf{q}}\partial_{\sigma_{h}(\kappa)\alpha}\hat{V}^{\mathrm{KS}}.
\end{equation}
Since conjugation by $\hat\sigma_h$ is linear and the prefactors are
c-numbers, the left-hand side is $\hat{\sigma}_{h}\Delta_{\mathbf{q}\nu}\hat{V}^{\mathrm{KS}}\hat{\sigma}_{h}^{-1}$,
and therefore
\begin{equation}
\hat{\sigma}_{h}\Delta_{\mathbf{q},\nu}\hat{V}^{\mathrm{KS}}\hat{\sigma}_{h}^{-1} = \bigg(\dfrac{\hbar}{2\omega_{\mathbf{q\nu}}}\bigg)^{1/2}\sum_{\kappa\alpha}\dfrac{\eta_{\alpha}e_{\kappa\alpha,\nu}(\mathbf{q})}{\sqrt{M_{\kappa}}}\Delta_{\mathbf{q}}\partial_{\sigma_{h}(\kappa)\alpha}\hat{V}^{\mathrm{KS}}.
\end{equation}
We make the variable substitution $\kappa'=\sigma_{h}(\kappa)$.  Since $\sigma_{h}$ is a bijection (each atom has a unique mirror image), if $\kappa$ runs through all the atoms, so does $\kappa'$. Thus Eq.~\eqref{eq:dV_both_sides} can be rewritten as,
\begin{equation}
\label{eq:dV_both_sides_rw}
\hat{\sigma}_{h}\Delta_{\mathbf{q},\nu}\hat{V}^{\mathrm{KS}}\hat{\sigma}_{h}^{-1} = \bigg(\dfrac{\hbar}{2\omega_{\mathbf{q\nu}}}\bigg)^{1/2}\sum_{\kappa'\alpha}\dfrac{\eta_{\alpha}e_{\sigma_{h}(\kappa')\alpha,\nu}(\mathbf{q})}{\sqrt{M_{\kappa'}}}\Delta_{\mathbf{q}}\partial_{\kappa'\alpha}\hat{V}^{\mathrm{KS}}.
\end{equation}
The action of $\hat\sigma_h$ on a polarization vector combines the
Cartesian sign change $\eta_\alpha$ with the atomic permutation
$\kappa'\to\sigma_h(\kappa')$. Since the lattice is symmetric under
$\hat\sigma_h$, each phonon mode has a well-defined parity
$\chi_\nu(\mathbf{q})=\pm1$, so that
$\eta_\alpha e_{\sigma_h(\kappa')\alpha,\nu}(\mathbf{q})
=\chi_\nu(\mathbf{q})\,e_{\kappa'\alpha,\nu}(\mathbf{q})$.
Thus, Eq.~\eqref{eq:dV_both_sides_rw} reads,
\begin{equation}
\hat{\sigma}_{h}\Delta_{\mathbf{q},\nu}\hat{V}^{\mathrm{KS}}\hat{\sigma}_{h}^{-1} = \bigg(\dfrac{\hbar}{2\omega_{\mathbf{q\nu}}}\bigg)^{1/2}\sum_{\kappa'\alpha}\dfrac{\chi_{\nu}(\mathbf{q})e_{\kappa'\alpha,\nu}(\mathbf{q})}{\sqrt{M_{\kappa'}}}\Delta_{\mathbf{q}}\partial_{\kappa'\alpha}\hat{V}^{\mathrm{KS}},
\end{equation}
and, factoring out $\chi_\nu(\mathbf{q})$ and comparing with
Eq.~\eqref{eq:Delta_qv},
\begin{equation}
\hat{\sigma}_{h}\Delta_{\mathbf{q},\nu}\hat{V}^{\mathrm{KS}}\hat{\sigma}_{h}^{-1} = \chi_{\nu}(\mathbf{q})\Delta_{\mathbf{q},\nu}\hat{V}^{\mathrm{KS}}.
\label{eq:chi}
\end{equation}
Therefore, the perturbation $\Delta_{\mathbf{q}\nu}\hat{V}^{\rm KS}$ inherits
the mirror parity $\chi_\nu(\mathbf{q})$ of the phonon mode. Inserting the
Bloch factorization, Eq.~\eqref{eq:bloch_factorization}, on both sides of
Eq.~\eqref{eq:chi}, and noting that $e^{i\mathbf{q}\cdot\mathbf{r}}$ is
mirror-even since $q_z=0$ in a 2D system, we find that the lattice-periodic
part of the perturbation carries the same parity,
\begin{equation}
\label{eq:v_chi}
\hat{\sigma}_{h}\Delta_{\mathbf{q}\nu}\hat{v}^{\rm KS}\hat{\sigma}_{h}^{-1}
=\chi_{\nu}(\mathbf{q})\,\Delta_{\mathbf{q}\nu}\hat{v}^{\rm KS}.
\end{equation}
On the other hand, the electronic Bloch states can be chosen as eigenstates
of the mirror operator,
\begin{equation}
\hat{\sigma}_h\psi_{n\mathbf{k}}(\mathbf{r})
=\psi_{n\mathbf{k}}(\sigma_h\mathbf{r})
=s_n(\mathbf{k})\,\psi_{n\mathbf{k}}(\mathbf{r}),
\qquad
s_n(\mathbf{k})=\pm 1.
\end{equation}
For a two-dimensional crystal, $\mathbf{k}=(k_x,k_y,0)$, and the horizontal
mirror leaves the Bloch phase unchanged,
$\mathbf{k}\cdot\sigma_h\mathbf{r}=\mathbf{k}\cdot\mathbf{r}$. Therefore,
using Eq.~\eqref{eq:bloch}, the periodic part of the Bloch state carries the
same mirror eigenvalue as the full Bloch state,
\begin{equation}
u_{n\mathbf{k}}(\sigma_h\mathbf{r})
=
s_n(\mathbf{k})\,u_{n\mathbf{k}}(\mathbf{r}).
\label{eq:u_sigma}
\end{equation}

To find the restrictions that mirror symmetry imposes on the matrix elements
$g_{mn\nu}(\mathbf{k},\mathbf{q})$, we start from
Eq.~\eqref{eq:g_matrix_element},
\begin{equation}
g_{mn\nu}(\mathbf{k},\mathbf{q})
= \langle u_{m\mathbf{k}+\mathbf{q}}|\Delta_{\mathbf{q}\nu}\hat{v}^{\mathrm{KS}}|u_{n\mathbf{k}}\rangle_{uc},
\end{equation}
and use the unitarity of $\hat{\sigma}_h$
($\hat{\sigma}_h^\dagger\hat{\sigma}_h=\hat{1}$), along with
Eqs.~\eqref{eq:v_chi} and \eqref{eq:u_sigma}:
\begin{equation}
\begin{split}
g_{mn\nu}(\mathbf{k},\mathbf{q})
&= \langle u_{m\mathbf{k}+\mathbf{q}}|\hat{\sigma}_{h}^{\dagger}\hat{\sigma}_{h}\,
\Delta_{\mathbf{q}\nu}\hat{v}^{\mathrm{KS}}\,
\hat{\sigma}_{h}^{\dagger}\hat{\sigma}_{h}|u_{n\mathbf{k}}\rangle_{uc}, \\
&= \langle \hat{\sigma}_{h}u_{m\mathbf{k}+\mathbf{q}}|\,
\hat{\sigma}_{h}\Delta_{\mathbf{q}\nu}\hat{v}^{\mathrm{KS}}\hat{\sigma}_{h}^{\dagger}\,
|\hat{\sigma}_{h}u_{n\mathbf{k}}\rangle_{uc}, \\
&= s_{m}(\mathbf{k}+\mathbf{q})\,\chi_{\nu}(\mathbf{q})\,s_{n}(\mathbf{k})\,
\langle u_{m\mathbf{k}+\mathbf{q}}|\Delta_{\mathbf{q}\nu}\hat{v}^{\mathrm{KS}}|u_{n\mathbf{k}}\rangle_{uc},
\end{split}
\end{equation}
i.e.,
\begin{equation}
g_{mn\nu}(\mathbf{k},\mathbf{q}) = \lambda_{mn\nu}(\mathbf{k},\mathbf{q})\,g_{mn\nu}(\mathbf{k},\mathbf{q}),
\end{equation}
where $\lambda_{mn\nu}(\mathbf{k},\mathbf{q})\equiv s_m(\mathbf{k}+\mathbf{q})\,\chi_\nu(\mathbf{q})\,s_n(\mathbf{k})=\pm1$.
Hence $g_{mn\nu}=-g_{mn\nu}=0$ whenever $\lambda_{mn\nu}=-1$, whereas for $\lambda_{mn\nu}=+1$ the relation is trivially satisfied. Therefore,
\begin{equation}
\label{eq:rule}
g_{mn\nu}(\mathbf{k},\mathbf{q}) = 0 \quad\text{unless}\quad \lambda_{mn\nu}(\mathbf{k},\mathbf{q}) = +1.
\end{equation}
It is worth noting that same result can be phrased in the language commonly used in chemistry and in the group theory of solids: a matrix element $\langle f|\hat O|i\rangle$ can be nonzero only if the direct product $\Gamma_f^{*}\otimes\Gamma_{\hat O}\otimes\Gamma_i$ contains the totally symmetric irreducible representation of the symmetry group common to
the initial state, the operator, and the final state~\cite{bradley1972mathematical,dresselhaus2008application,
cotton1990group}. For the e-ph matrix element, $\Gamma_i$, $\Gamma_f$, and $\Gamma_{\hat O}$ are the irreducible representations of $\psi_{n\mathbf k}$, $\psi_{m,\mathbf k+\mathbf q}$, and the deformation potential $\Delta_{\mathbf{q},\nu}v^{\mathrm{KS}}(\mathbf{r})$, the last being that of the phonon mode $\nu$ (Eq.~\eqref{eq:v_chi}). In a two-dimensional crystal, the only symmetry element guaranteed to leave $\mathbf k$, $\mathbf q$, and $\mathbf k+\mathbf q$ invariant at every point of the Brillouin zone is $\sigma_h$, so the relevant point group is  $C_s=\{E,\sigma_h\}$. Its irreducible representations are $A'$  (character $+1$ under $\sigma_h$) and $A''$ (character $-1$), with $A'\otimes A'=A''\otimes A''=A'$ and $A'\otimes A''=A''$. The triple product therefore contains the totally symmetric representation $A'$
exactly when $s_m(\mathbf k+\mathbf q)\,\chi_\nu(\mathbf q)\,
s_n(\mathbf k)=+1$, which is Eq.~\eqref{eq:rule}.
This formulation also highlights the generality of Eq.~\eqref{eq:rule}: it is the symmetry constraint shared by every point of the two-dimensional Brillouin zone. At high-symmetry points, where the group
common to $\mathbf k$, $\mathbf q$, and $\mathbf k+\mathbf q$ is larger than $C_s$, it is complemented by additional, point-specific constraints
that can impose further zeros. In this sense, Eq.~\eqref{eq:rule} is a necessary condition, rather than the complete set of symmetry constraints at every wave vector. At high-symmetry points
the group common to $\mathbf k$, $\mathbf q$, and $\mathbf k+\mathbf q$  is larger than $C_s$ (for instance $D_{3h}$ at $\Gamma$ and $C_{2v}$ at
M for the $P\bar6m2$ crystal studied below), and the requirement that $\Gamma_f^{*}\otimes\Gamma_\nu\otimes\Gamma_i$ contain the totally
symmetric representation of that larger group can impose additional zeros. For example, at $\mathbf k=\mathbf q=\Gamma$ the intraband matrix element of a nondegenerate band can be nonzero only for totally  symmetric ($A_1'$) phonons, since $\Gamma_n^{*}\otimes\Gamma_n=A_1'$ for any one-dimensional $\Gamma_n$.

Table~\ref{tab:selection_rule} summarizes the resulting selection rule for 
all combinations of electronic mirror parity ($s_n,s_m=\pm1$) and phonon 
mode parity ($\chi_\nu=\pm1$). Allowed transitions are those satisfying 
$s_n(\mathbf{k})\,\chi_\nu(\mathbf{q})\,s_m(\mathbf{k}+\mathbf{q})=+1$. 
Physically, even phonons ($\chi_\nu=+1$) preserve the electronic mirror 
parity, allowing only $\sigma\to\sigma$ and $\pi\to\pi$ intraband 
processes, whereas odd phonons ($\chi_\nu=-1$) flip it, allowing 
$\sigma\to\pi$ and $\pi\to\sigma$ interband processes. This rule holds 
throughout the entire Brillouin zone of a two-dimensional crystal, as 
$\sigma_h$ belongs to the little group of every in-plane wave vector.

\begin{table}[h]
\caption{\label{tab:selection_rule}Mirror-parity selection rule for
the electron--phonon matrix element $g_{mn,\nu}(\mathbf k,\mathbf q)$.
Electronic states of mirror eigenvalue $s=+1$ ($\sigma$) and $s=-1$
($\pi$) couple through phonon modes of parity $\chi_\nu=+1$ (even) or
$\chi_\nu=-1$ (odd) only when $s_n(\mathbf k)\,\chi_\nu(\mathbf q)\,
s_m(\mathbf k+\mathbf q)=+1$.}
\begin{ruledtabular}
\begin{tabular}{cccc}
$s_n(\mathbf k)$ & $s_m(\mathbf k+\mathbf q)$ & $\chi_\nu(\mathbf q)$ & Coupling \\
\hline
$\sigma$ & $\sigma$ & even & allowed ($\sigma\sigma$) \\
$\sigma$ & $\sigma$ & odd  & forbidden \\
$\pi$    & $\pi$    & even & allowed ($\pi\pi$) \\
$\pi$    & $\pi$    & odd  & forbidden \\
$\sigma$ & $\pi$    & even & forbidden \\
$\sigma$ & $\pi$    & odd  & allowed ($\sigma\pi$) \\
$\pi$    & $\sigma$ & even & forbidden \\
$\pi$    & $\sigma$ & odd  & allowed ($\pi\sigma$) \\
\end{tabular}
\end{ruledtabular}
\end{table}

\section{Numerical examples}
To illustrate the implications of the electron--phonon selection rule,
we present DFPT calculations for two representative systems with
horizontal mirror symmetry, $\sigma_h$. We first consider
\ce{Sc2BCH4}, an hypothetical monolayer derived from the recently proposed family of
transition-metal borides identified as promising two-dimensional
conventional superconductors
\cite{sbbm-hvc4,seeyangnok2025high}. In this strictly two-dimensional
system, $\sigma_h$ belongs to the little group of every electronic and
phononic wave vector, allowing us to test the selection rule throughout
the entire Brillouin zone. We then analyze bulk \ce{MgB2}, an
archetypal multiband conventional superconductor, to investigate how
the same symmetry constraints operate in three dimensions. In this
case, mirror parity is well defined only on mirror-invariant subspaces
of the Brillouin zone, enabling us to examine both the exact selection
rule on these subspaces and the evolution of the electron--phonon
matrix elements as the electronic states move away from them.

\begin{figure*}
\centering
\includegraphics[scale=0.75]{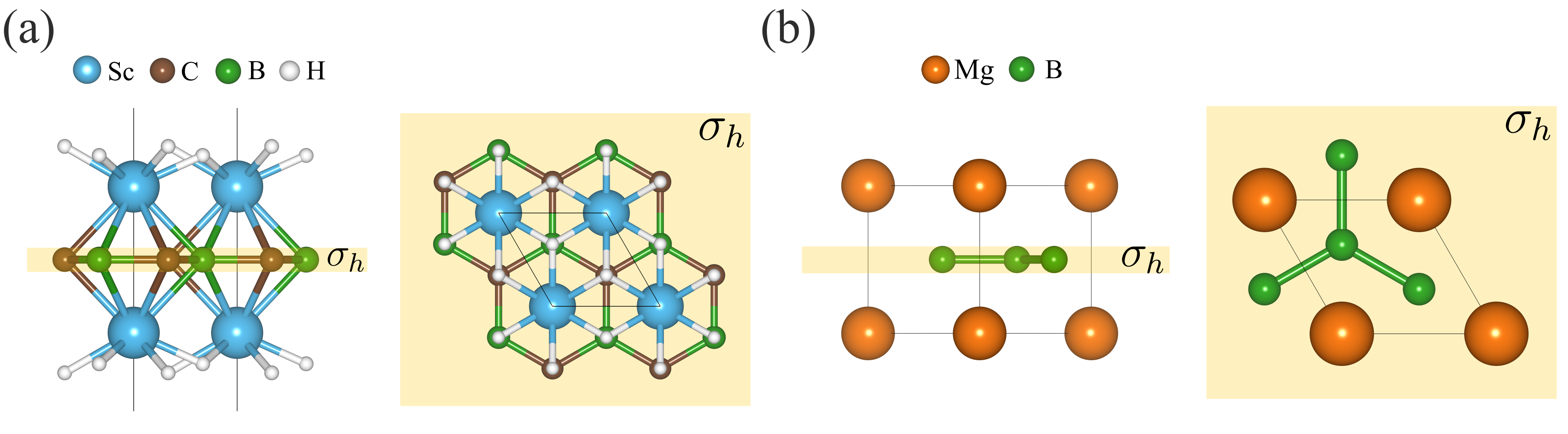}
\caption{Top and side views of the optimized crystal structures of
(a) \ce{Sc2BCH4} and (b) \ce{MgB2}. The corresponding primitive unit
cells are indicated by black solid lines.}
\label{fig:Crystal_structures}
\end{figure*}

\subsection{Electron-phonon coupling and superconducting properties of \ce{Sc2BCH4}}
The  crystal structure of \ce{Sc2BCH4} has hexagonal symmetry and belongs to the space group \(P\bar{6}m2\) (No.~187). We report the optimized lattice parameters and Wyckoff positions in the  Supplemental Materials (SM). Figure~\ref{fig:Crystal_structures}(a) shows top and side views of the crystal structure. The monolayer has a horizontal mirror plane, \(\sigma_h\), located within the BC layer and perpendicular to the nonperiodic direction, which relates the two sides of the monolayer. As we already mentioned, because $\sigma_h$ leaves every in-plane $\mathbf{k}$ and $\mathbf{q}$ vector invariant, electronic states and phonon modes can be labeled by their mirror parity throughout the entire Brillouin zone.

\begin{figure*}
\centering
\includegraphics[scale=0.3]{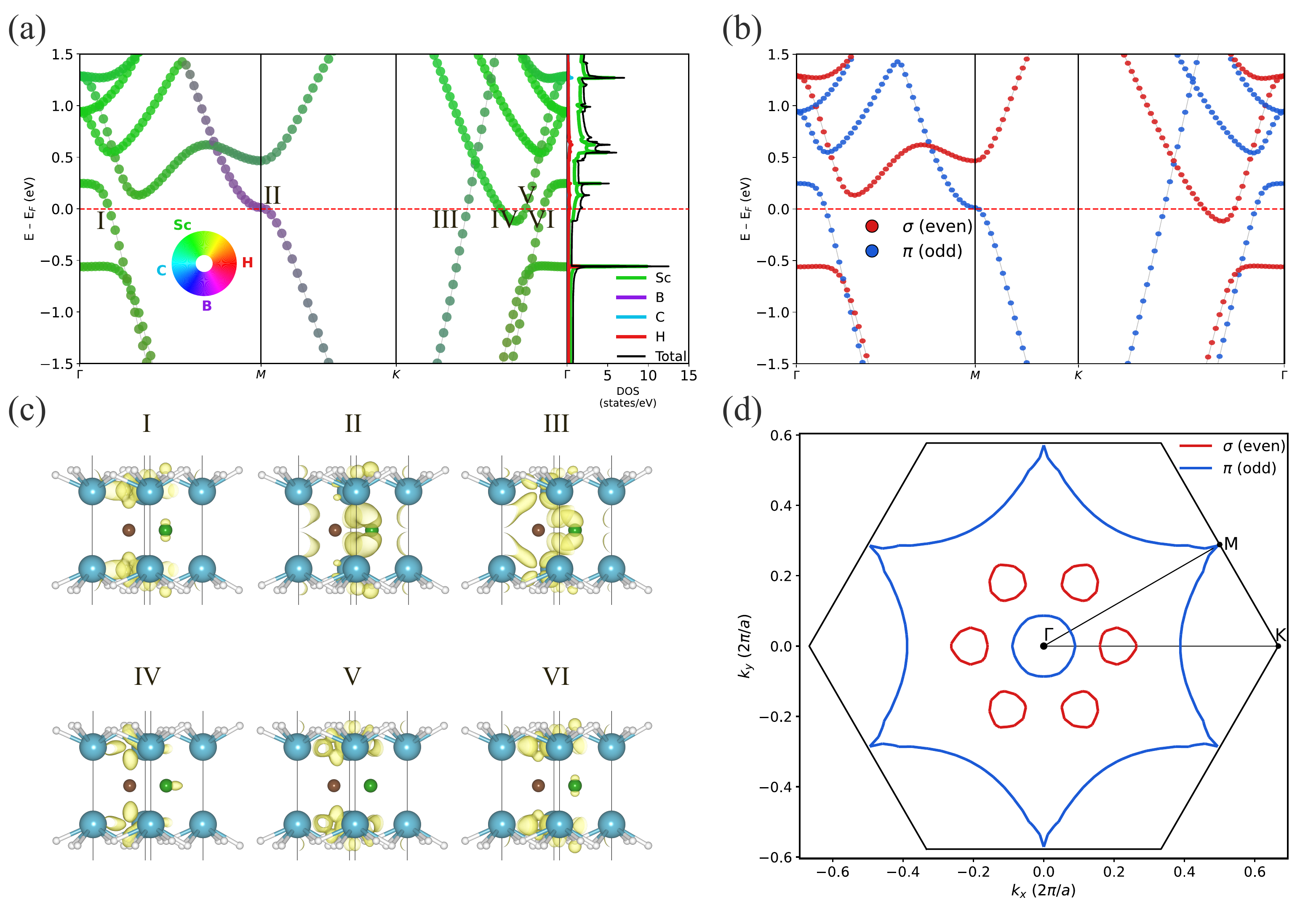}
\caption{Electronic structure of \ce{Sc2BCH4}. (a) Atom-resolved band structure,
with the color of the markers indicating the projection of each
electronic state onto the constituent atoms. The side panels show the corresponding projected electronic
density of states (DOS). (b) Band structure resolved
according to the eigenvalue of the horizontal mirror $\sigma_h$:
mirror-even ($\sigma$) states in red and mirror-odd ($\pi$) states in
blue. (c) Real-space probability densities
$|\psi_{n\mathbf{k}}(\mathbf{r})|^2$ of six Kohn--Sham states crossing
the Fermi level, labeled I--VI in (b). (d) Two-dimensional Fermi surface
colored according to the mirror parity of the electronic states.}
\label{fig:bands_sigma_pi_Sc2BCH4_2}
\end{figure*}

We characterize the electronic structure of \ce{Sc2BCH4} by calculating its electronic band structure, density of states (DOS), and Fermi surface within DFT (see Sec.~\ref{sec:computational} for computational details). Figure~\ref{fig:bands_sigma_pi_Sc2BCH4_2}(a) shows the electronic band structure within an energy window of $\pm 1.5$ eV around the Fermi level, projected onto the  atomic species. Three bands cross the Fermi level multiple times along the high-symmetry path, revealing the metallic character of \ce{Sc2BCH4}. Among these Fermi-level crossings, one feature stands out. Around $\mathbf{k}=\mathrm{M}$,  a van Hove singularity lies at the Fermi level, giving rise to a subtle increase in the electronic DOS. Van Hove singularities near the Fermi level have been associated with enhanced electron--phonon coupling, as the increased electronic density of states can enlarge  the phase space available for electron--phonon scattering\cite{yang2023theoretical,liu2024three,gai2022van,meng2025mbenes,meng2024unraveling,xie2025hydrogenation,kim2018importance,sano2016effect,quan2016van,chen2024high,seeyangnok2025high}. In contrast, the remaining Fermi-level crossings involve more dispersive bands with relatively large slopes and large projection onto Sc atoms. There is a relevant nesting in the different Fermi surfaces. Panel (c) of Fig.~\ref{fig:bands_sigma_pi_Sc2BCH4_2} shows the real-space probability densities, $|\psi_{n\mathbf{k}}(\mathbf{r})|^2$, of six particular Kohn--Sham states, $|\psi_{n,\mathbf{k}}\rangle$,  crossing the Fermi level, providing a direct view of their spatial character.

The presence of the horizontal mirror plane allows us to classify every electronic state according to its parity under $\sigma_h$. We therefore label the mirror-even states as $\sigma$ and the mirror-odd states as $\pi$, following the usual chemical convention for finite systems\cite{mcnaught1997compendium}. Figure~\ref{fig:bands_sigma_pi_Sc2BCH4_2}(b) shows the electronic band structure colored according to this mirror character. As can be seen, four of the six selected states crossing the Fermi level have $\pi$ character, whereas the remaining two correspond to $\sigma$ states originating from the same electronic band. The corresponding two-dimensional Fermi surface is shown in Fig.~\ref{fig:bands_sigma_pi_Sc2BCH4_2}(d), with the Fermi contours colored according to their mirror parity. The $\sigma$ states form six symmetry-related, nearly circular Fermi pockets, which are intersect by the $\Gamma-\mathrm{K}$ high-symmetry path. The $\pi$ states, in contrast, give rise to a small circular pocket centered at $\Gamma$, intersecting both the $\Gamma-\mathrm{K}$ and $\Gamma-\mathrm{M}$ directions, together with a larger,  warped sixfold-symmetric Fermi contour.

\begin{figure*}
\centering
\includegraphics[scale=0.50]{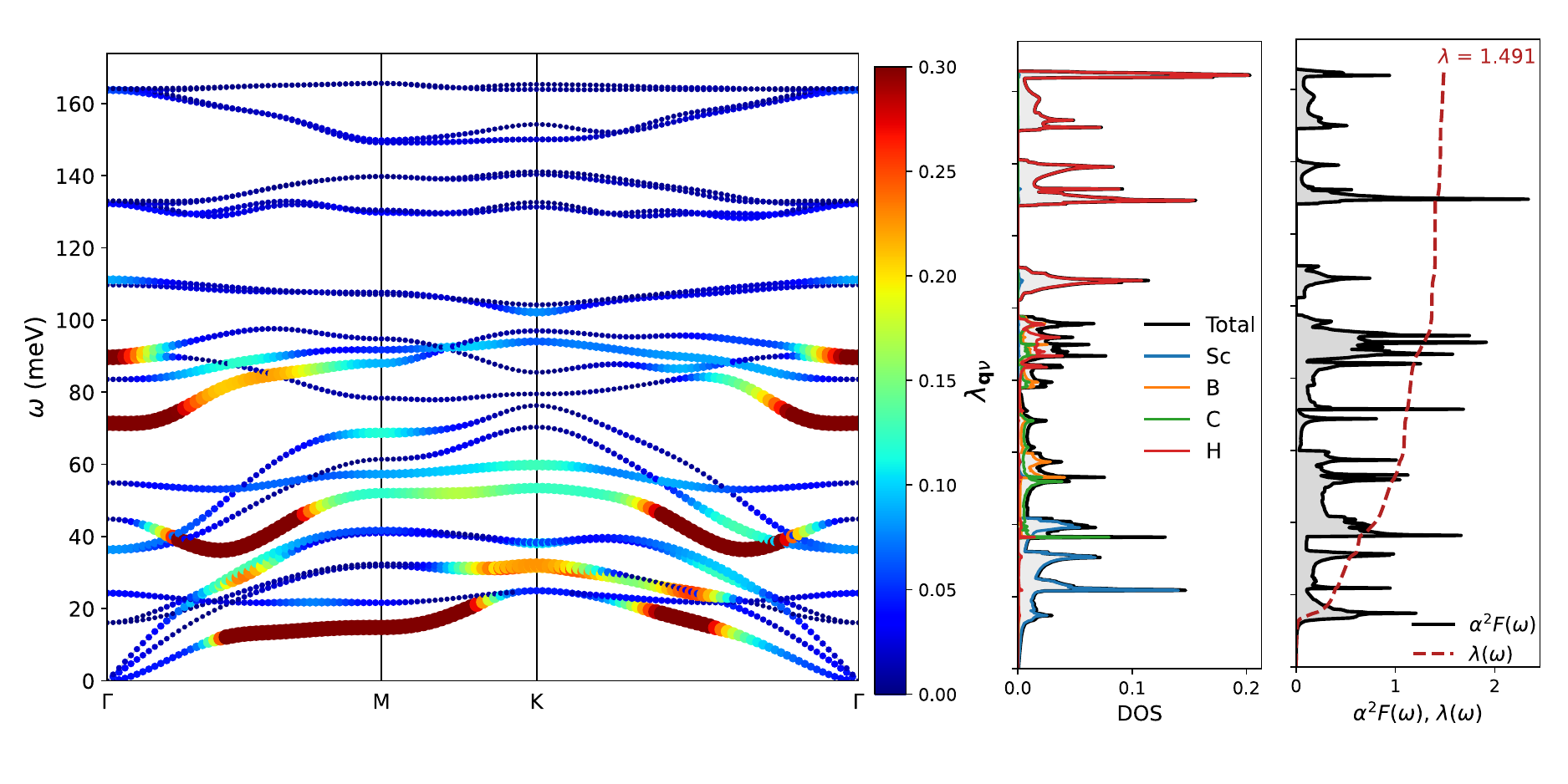}
\caption{Phonon dispersions colored by the mode-resolved electron-phonon coupling strength $\lambda_{\mathbf{q}\nu}$, atom-projected phonon density of states, and Eliashberg spectral function $\alpha^{2}F(\omega)$ together with the cumulative electron-phonon coupling $\lambda(\omega)$ for \ce{Sc2BCH4}.}
\label{fig:Phonon_dispersion_lambda_DOS_Eliashberg}
\end{figure*}

The dynamical stability, electron--phonon coupling, and superconducting properties of \ce{Sc2BCH4} were evaluated within DFPT (see Sec.~\ref{sec:computational} for computational details). Figure~\ref{fig:Phonon_dispersion_lambda_DOS_Eliashberg} shows its phonon dispersion. The absence of imaginary-frequency modes throughout the Brillouin zone confirms the dynamical stability of the monolayer. The color-scale represents the mode($\nu$)- and momentum($\mathbf q$)-resolved electron--phonon coupling strength, $\lambda_{\mathbf q\nu}$. The selection rule derived above allows us to interpret the distribution of the electron--phonon coupling in terms of the mirror parities of both the phonon modes and the electronic states involved in the scattering processes.

We first focus on the zone center, $\mathbf q=\Gamma$. where the electron--phonon matrix element  reduces exactly to $g_{mn,\nu}(\mathbf k,\Gamma)
=
\left\langle
u_{m\mathbf k}
\left|
\Delta_{\Gamma\nu}v^{\mathrm{KS}}
\right|
u_{n\mathbf k}
\right\rangle$, so that both intra- and interband scattering processes conserve the
the electronic quasimomentum. As shown in Fig.~\ref{fig:Phonon_dispersion_lambda_DOS_Eliashberg} and Table~\ref{tab:modes_channels}, three phonon modes exhibit significant electron--phonon coupling at $\mathbf{q}=\Gamma$: the nondegenerate mode $\nu=11$ and the doubly degenerate pair $\nu=13,14$. Mode $\nu=11$ is dominated by out-of-plane displacements of the H atoms located on both sides of the  layer, whereas the degenerate modes $\nu=13$ and $14$ predominantly involve in-plane displacements of both
the H atoms and the atoms forming the honeycomb BC layer (see Fig. S2 in SM).

Since superconductivity is governed primarily by scattering between electronic states at the Fermi surface, we restrict the analysis to these processes. To this end, we define the Fermi-surface-restricted electron--phonon scattering weight as
\begin{equation}
P_{\nu}(\mathbf{k},\mathbf{q})
=
\sum_{mn}
|g_{mn,\nu}(\mathbf{k},\mathbf{q})|^{2}
\delta_{\sigma}(\varepsilon_{n\mathbf{k}}-\varepsilon_F)
\delta_{\sigma}(\varepsilon_{m,\mathbf{k}+\mathbf{q}}-\varepsilon_F).
\end{equation}
This quantity measures the squared electron--phonon matrix elements associated with a given $(\mathbf k,\mathbf q)$ pair, weighted so as to select initial and final states near the Fermi surface.  For computational purposes the Dirac  delta is represented with  a Gaussian function with a broadening of $\sigma=0.005$~Ry.

\begin{figure*}
\centering
\includegraphics[scale=0.30]{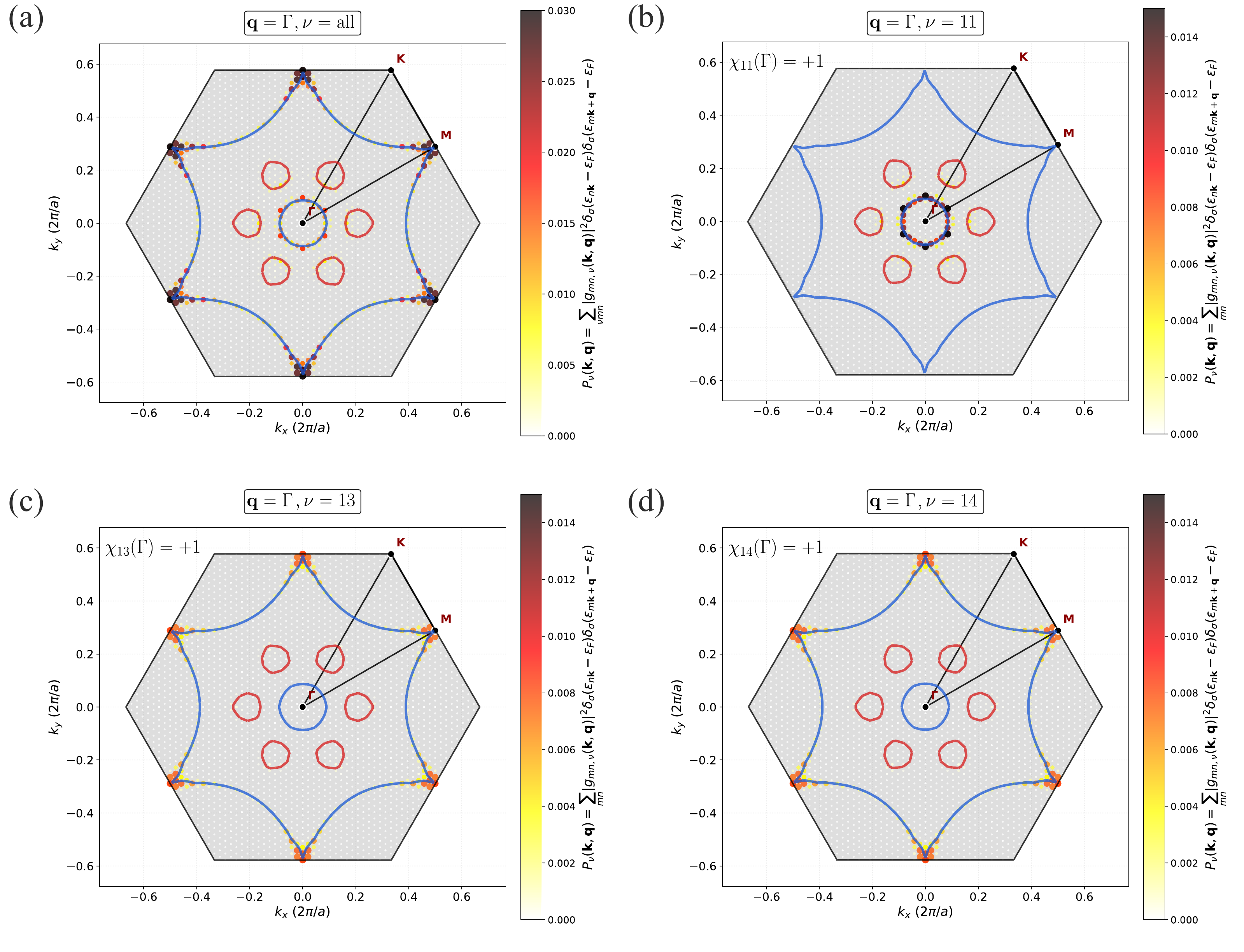}
\caption{Fermi-surface-restricted electron--phonon scattering weight
$P_{\nu}(\mathbf{k},\mathbf{q})=
\sum_{mn}|g_{mn,\nu}(\mathbf{k},\mathbf{q})|^{2}
\delta_{\sigma}(\varepsilon_{n\mathbf{k}}-\varepsilon_F)
\delta_{\sigma}(\varepsilon_{m,\mathbf{k}+\mathbf{q}}-\varepsilon_F)$
at $\mathbf{q}=\Gamma$, using a Gaussian broadening of
$\sigma=0.005$~Ry. The Fermi surface, colored according to the
electronic mirror parity under $\sigma_h$ (even:red ; odd: blue),
is superimposed on each panel. The dots represent the
electron--phonon scattering processes and are colored according to
the magnitude of the corresponding scattering weight.
(a) Total scattering weight summed over all phonon branches,
$\sum_{\nu}P_{\nu}$.
(b) Contribution from phonon mode $\nu=11$.
(c,d) Contributions from the degenerate phonon modes $\nu=13$ and
$\nu=14$, respectively.}
\label{fig:bz_kpoints_mode_all_11_13_14}
\end{figure*}

Figure~\ref{fig:bz_kpoints_mode_all_11_13_14}(a) shows the total Fermi-surface-restricted scattering weight,
\begin{equation}
P(\mathbf k,\mathbf q)
=
\sum_{\nu}
P_{\nu}(\mathbf k,\mathbf q),
\end{equation}
with the Fermi surface superimposed and colored according to its mirror parity. The dominant contributions appear as dark spots and are located almost entirely on the
$\pi$-parity Fermi contours, indicating that the electron--phonon coupling is dominated by intraband $\pi\rightarrow\pi$ scattering. According to the selection rule, $\pi\rightarrow\pi$ scattering is allowed
only through mirror-even ($\sigma$) phonons. Consistently, modes $\nu=11$, $13$, and $14$ are all even under $\sigma_h$ (see Table~S6 in the Supplemental Material).

\begin{table*}
\caption{Phonon modes of \ce{Sc2BCH4} that
allow each electronic scattering channel at $\mathbf q=\Gamma$ (point
group $D_{3h}$) and $\mathbf q=\mathrm{M}$ (point group of the wave vector $C_{2v}$).
Modes are grouped by irreducible representation; mode indices $\nu$ and irreducible representations follow
Tables~S6 and S7 of the Supplemental Material; acoustic modes at
$\Gamma$ are marked with an asterisk. The last column lists the modes
with the largest mode-resolved coupling $\lambda_{\mathbf q\nu}$ in each
channel.}
\label{tab:modes_channels}
\begin{ruledtabular}
\begin{tabular}{c|c|c|c|c}
Channel & $\chi_\nu(\mathbf q)$ & $\mathbf q$ & Allowed modes $\nu$ & Strongest $\nu$ ($\lambda_{\mathbf q\nu}$) \\
\hline
$\sigma\to\sigma$, $\pi\to\pi$ & $+1$ & $\Gamma$ &
  $A_1'$: 6, 11, 16;\; $E'$: 1--2$^{*}$, 7--8, 13--14, 17--18, 21--22 &
  11 (0.45), 13--14 (0.32) \\
 & & M &
  $A_1$: 2, 6, 10, 12, 14, 15, 17, 22;\; $B_1$: 3, 9, 11, 20, 23 &
  10 (0.12), 12 (0.09) \\
$\sigma\to\pi$, $\pi\to\sigma$ & $-1$ & $\Gamma$ &
  $A_2''$: 3$^{*}$, 9, 10, 12, 15;\; $E''$: 4--5, 19--20, 23--24 &
  all $\lambda_{\Gamma\nu}<0.01$ \\
 & & M &
  $A_2$: 4, 19, 24;\; $B_2$: 1, 5, 7, 8, 13, 16, 18, 21 &
  1 (0.49), 7 (0.12) \\
\end{tabular}
\end{ruledtabular}
\end{table*}

Figures~\ref{fig:bz_kpoints_mode_all_11_13_14}(b)--(d) resolve $P(\mathbf k,\Gamma)$ into the contributions
from these three modes. Mode $\nu=11$ predominantly couples the $\pi$
states forming the circular Fermi pocket centered at $\Gamma$, whereas
the degenerate modes $\nu=13$ and $14$ couple mainly the $\pi$ states
near the $M$ points, associated with the flat band and the corresponding
van Hove singularity. 

It is worth noting that several other phonon modes also satisfy the symmetry condition for these scattering processes. Figure~S3
shows $P_{\nu}(\mathbf{k},\Gamma)$ for all 24 phonon modes.  For instance,
modes $\nu=7$, $8$, and $16$, which appear in light blue in
Fig.~\ref{fig:Phonon_dispersion_lambda_DOS_Eliashberg}, are symmetry allowed but
exhibit much weaker coupling. This illustrates an important distinction:
the selection rule determines whether an electron--phonon matrix element
is allowed to be nonzero, but it does not determine its magnitude.
A symmetry-forbidden process therefore yields a vanishing matrix
element, up to numerical noise in the DFPT calculation, whereas the
strength of a symmetry-allowed process depends on the detailed overlap
between the initial and final Kohn--Sham states and the phonon-induced
deformation potential. 

\begin{figure*}
\centering
\includegraphics[scale=0.475]{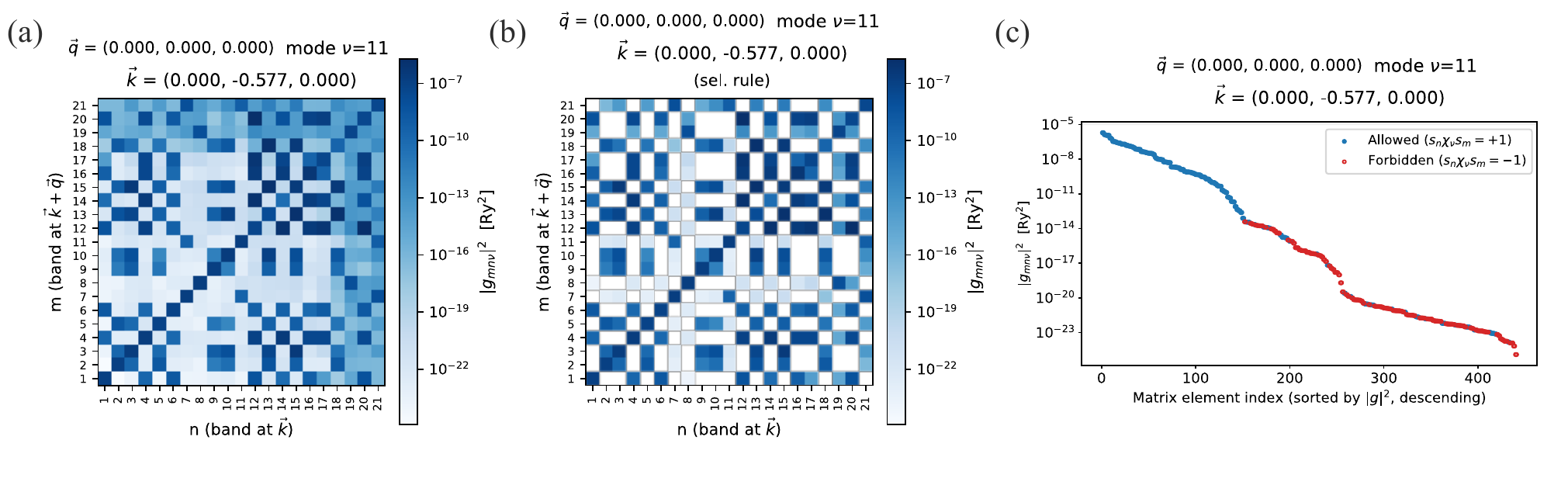}
\caption{Electron--phonon matrix elements and their symmetry filtering
for $\mathbf{q}=\Gamma$ and $\mathbf{k}=\mathrm{M}$. (a) Squared electron--phonon matrix elements,
$|g_{mn,\nu}(\mathbf k,\mathbf q)|^2$, with the color scale indicating
their magnitude. (b) Same matrix after explicitly setting to zero all
symmetry-forbidden elements, i.e.,
$s_n(\mathbf k)\chi_\nu(\mathbf q)s_m(\mathbf k+\mathbf q)=-1$.
(c) Matrix elements sorted in descending order of $|g_{mn,\nu}|^2$; symmetry-allowed
and symmetry-forbidden elements are shown in blue and red, respectively.}
\label{fig:G_matrix_symmetries_values}
\end{figure*}

We examine this point in greater detail in Fig.~\ref{fig:G_matrix_symmetries_values}. Panels (a) and (b) show the squared electron--phonon matrix elements, $|g_{mn,\nu}(\mathbf{k},\mathbf{q})|^2$, for $\mathbf{q}=\Gamma$, $\mathbf{k}=M'$, and phonon mode $\nu=11$, with the color scale representing their magnitude on a logarithmic scale. Panel (a) shows the matrix elements obtained directly from the QE DFPT calculation, whereas panel (b) shows the same matrix after setting all symmetry-forbidden elements identically to zero. Remarkably, this procedure removes no matrix element with an appreciable magnitude, leaving the overall coupling pattern essentially unchanged. This separation becomes even clearer in Fig.~\ref{fig:G_matrix_symmetries_values} (c), where the matrix elements from panel (a) are sorted in descending order of $|g|^2$ and colored according to the selection rule: blue for symmetry-allowed elements and red for symmetry-forbidden ones. The largest symmetry-forbidden matrix elements are approximately eight orders of magnitude smaller than the dominant allowed elements and can therefore be regarded as numerical noise within the DFPT calculation. At the same time, symmetry-allowed matrix elements span a broad range of magnitudes and may themselves be numerically negligible, further
emphasizing that satisfying the selection rule is a necessary, but not
sufficient, condition for strong electron--phonon coupling.

\begin{figure}
\centering
\includegraphics[scale=0.35]{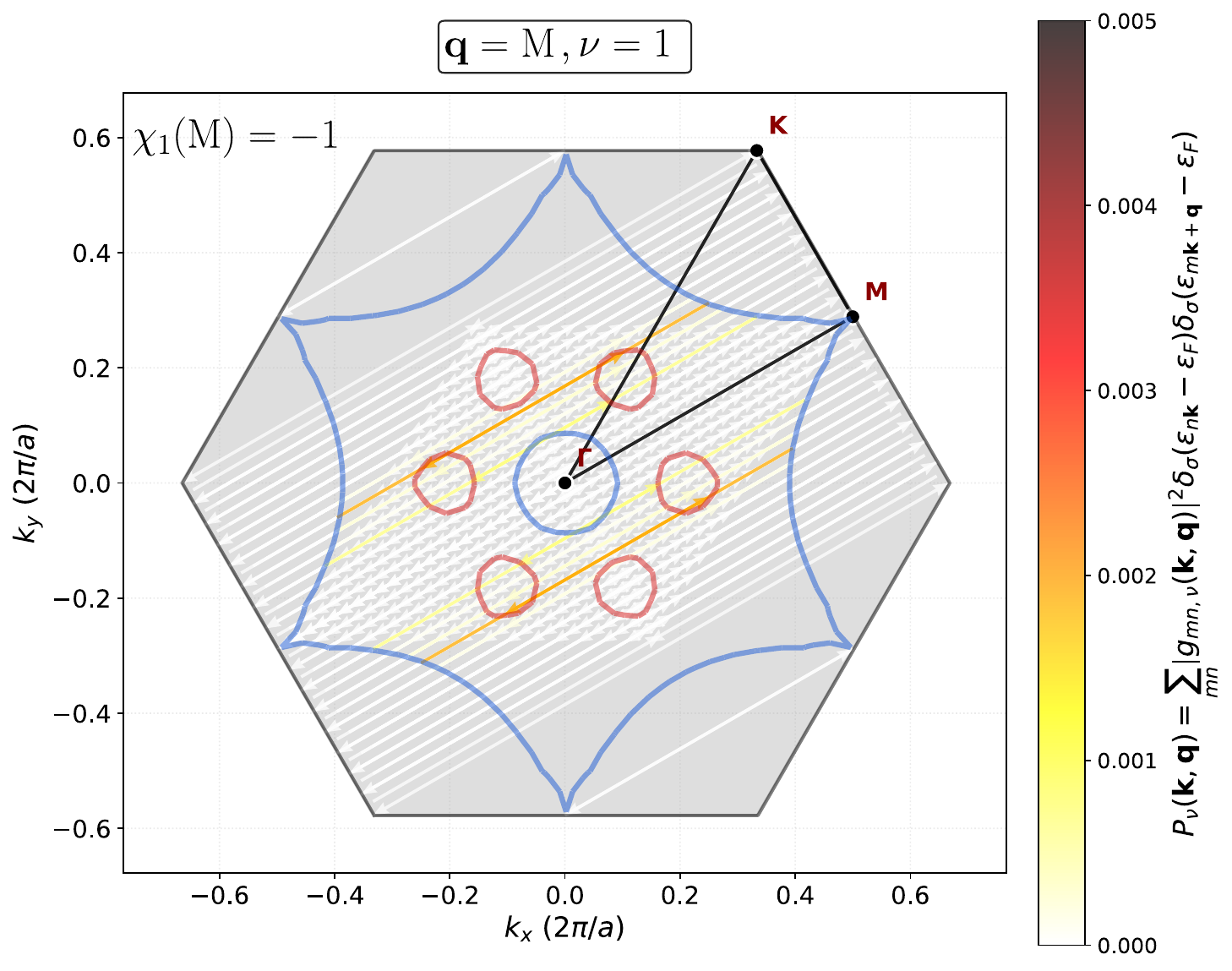}
\caption{Fermi-surface-restricted electron--phonon scattering weight
$P_{\nu}(\mathbf{k},\mathbf{q})=
\sum_{mn}|g_{mn,\nu}(\mathbf{k},\mathbf{q})|^{2}
\delta_{\sigma}(\varepsilon_{n\mathbf{k}}-\varepsilon_F)
\delta_{\sigma}(\varepsilon_{m,\mathbf{k}+\mathbf{q}}-\varepsilon_F)$
at $\mathbf{q}=\mathrm{M}$, using a Gaussian broadening of
$\sigma=0.005$~Ry. The Fermi surface, colored according to the
electronic mirror parity under $\sigma_h$ (even:red ; odd: blue),
is superimposed. The arrows show the vector $\bm{q}=M$ connecting the Fermi surfaces.}
\label{fig:bz_kpoints_mode_1_q_M}
\end{figure}

Having analyzed the $\mathbf{ q}=\Gamma$ case, we now turn to finite-$\mathbf{q}$ scattering, where the electron changes its quasimomentum from $\mathbf{k}$ to $\mathbf{k}+\mathbf{q}$. In this case, the Fermi-surface-restricted electron--phonon scattering weight can be visualized as arrows connecting the initial and final electronic states on the Fermi surface, with each arrow corresponding to a phonon-mediated transition carrying wave vector $\mathbf{q}$ and therefore having length $|\mathbf q|$ in reciprocal space.
For the particular case $\mathbf {q}=\mathrm{M}$, Fig.~\ref{fig:Phonon_dispersion_lambda_DOS_Eliashberg} reveals at least three phonon modes with appreciable electron--phonon coupling, the strongest contribution arising from the lowest-frequency mode.

Figure~\ref{fig:bz_kpoints_mode_1_q_M}(a) shows the Fermi-surface-restricted electron--phonon scattering weight for this lowest-frequency mode. The mode is odd under $\sigma_{h}$, and therefore the selection rule restricts its coupling to electronic states of opposite mirror parity. The dominant processes, highlighted by the warmest arrows, indeed connect $\pi$ states on the strongly warped sixfold Fermi contour with $\sigma$ states forming the nearly circular Fermi pockets. The remaining scattering channels carry substantially smaller weight and therefore make only a minor contribution to the total electron--phonon coupling for this mode.

Taken together, these results establish \ce{Sc2BCH4} as a clear
two-dimensional realization of the mirror-parity selection rule.
Because $\sigma_h$ belongs to the little group of every electronic and
phononic wave vector, the symmetry constraint applies throughout the
entire Brillouin zone. The $\mathbf q=\Gamma$ analysis shows that
mirror-even phonons mediate scattering between electronic states of the
same parity, while the finite-$\mathbf q$ example at
$\mathbf q=\mathrm{M}$ demonstrates that mirror-odd phonons couple
states of opposite parity. In both cases, symmetry-forbidden matrix
elements vanish up to numerical noise, whereas symmetry-allowed
elements may span a wide range of magnitudes. The selection rule
therefore acts as an exact symmetry filter for electron--phonon
scattering channels, while the strength of the allowed processes
remains determined by the microscopic electronic and vibrational
structure.

Beyond providing a clean realization of the mirror-parity selection rule, the hypothetical material 
\ce{Sc2BCH4} also exhibits strong phonon-mediated superconductivity. Our 
anisotropic Migdal--Eliashberg calculations yield a critical temperature of 
approximately 100~K, with the corresponding superconducting gap reported in Fig.~S5
the SI. These results establish \ce{Sc2BCH4} as a 
high-$T_c$ conventional superconductor within the present first-principles 
framework. We stress, however, that these results do not by themselves 
establish the experimental accessibility of the monolayer. While the harmonic 
phonon spectrum confirms dynamical stability, actual realization also depends 
on thermodynamic and kinetic stability, as well as on viable synthesis or 
stabilization pathways; a detailed assessment of these aspects lies beyond the 
scope of this work, which focuses on the symmetry constraints governing the 
electron--phonon interaction.

\subsection{Electron-phonon coupling and superconducting properties of \ce{MgB2}}

Bulk \ce{MgB2} is the conventional superconductor with the highest
experimentally measured critical temperature at ambient pressure,
$T_c\simeq39$~K~\cite{nagamatsu2001superconductivity,hinks2001complex,choi2002origin,souma2003origin}.
It crystallizes in the hexagonal AlB$_2$-type structure
with space group $P6/mmm$ (No.~191). The B atoms form planar
graphite-like honeycomb sheets stacked directly along the $c$ axis,
with no lateral displacement between consecutive layers, while the Mg
atoms occupy the interlayer region. The resulting structure can be
viewed as a stacking of hexagonal B layers separated by Mg planes. Figure~\ref{fig:Crystal_structures}(b) shows side and top views of the
crystal structure. As in \ce{Sc2BCH4}, \ce{MgB2} possesses a horizontal
mirror plane, $\sigma_h$, perpendicular to the principal rotation axis;
in particular, one such mirror plane coincides with the honeycomb B
layer.

\begin{figure*}
\centering
\includegraphics[scale=0.29]{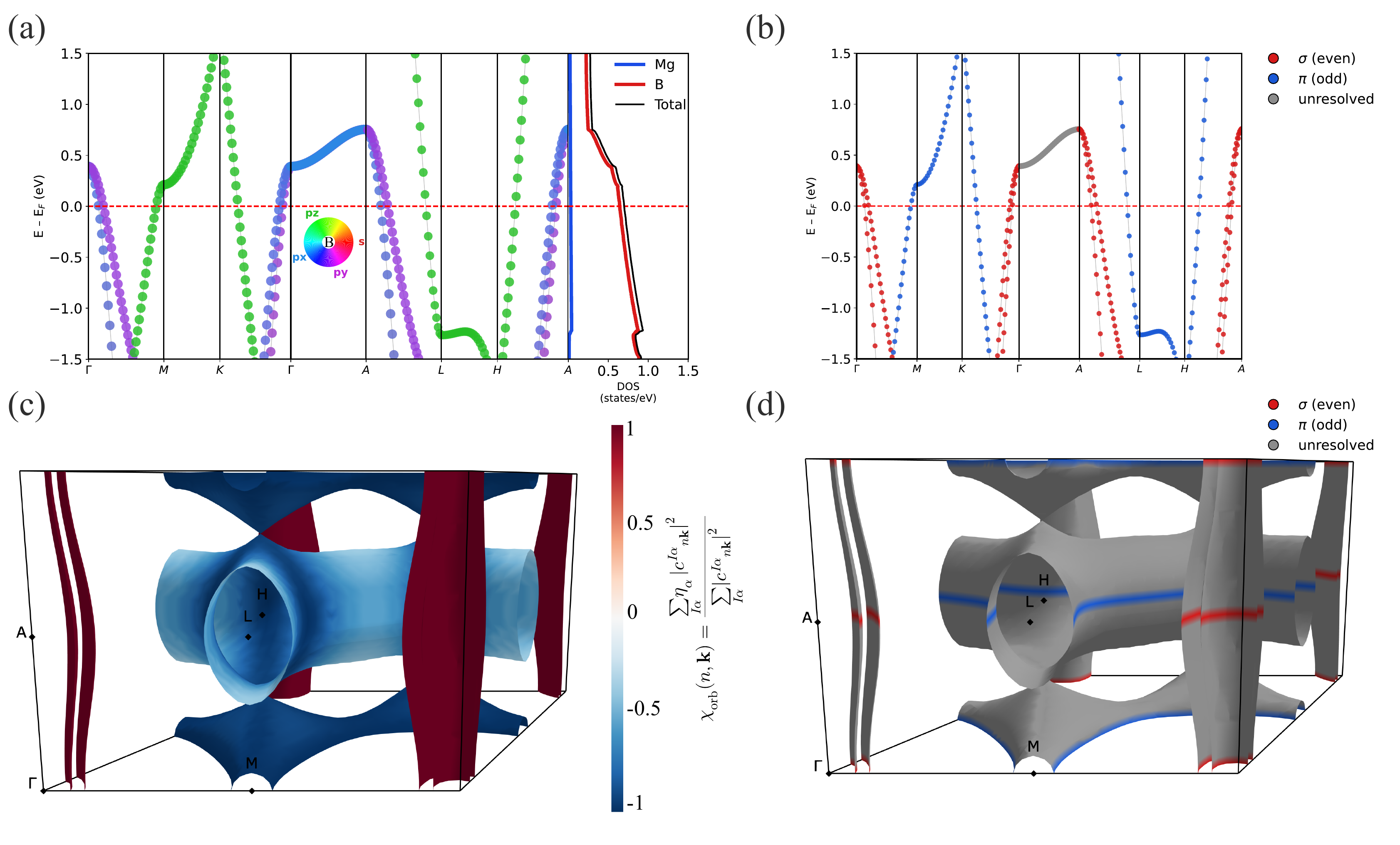}
\caption{Electronic structure and mirror character of \ce{MgB2}.
(a) Electronic band structure projected onto the B $s$, $p_x$, $p_y$,
and $p_z$ orbitals, together with the corresponding projected and total
electronic density of states (DOS).
(b) Electronic bands classified by their eigenvalue under the horizontal
mirror operation $\sigma_h$ wherever mirror parity is well defined. Three-dimensional Fermi surface colored by the orbital-character
indicator $\chi_\mathrm{orb}(\mathbf{k})$ [Eq.~(S4.13)],
the normalized difference between the projected weights on in-plane
($s$, $p_x$, $p_y$) and out-of-plane ($p_z$) atomic orbitals of all atoms. Red and blue denote states dominated by even and odd orbitals, respectively. 
(d) Three-dimensional Fermi surface showing the exact $\sigma_h$
eigenvalue on the mirror-invariant planes $k_z=0$ and $k_z=\pi/c$,
where electronic states possess a well-defined mirror parity.}
\label{fig:bands_FS_atoms_sigma_pi_MgB2}
\end{figure*}

There is, however, a fundamental difference between bulk \ce{MgB2}
and a strictly two-dimensional crystal with a plane $\sigma_h$.
In two dimensions, every electronic and phononic wave vector lies in
the mirror plane and is therefore left invariant by $\sigma_h$. In the
three-dimensional Brillouin zone of \ce{MgB2}, instead, a generic wave vector is not invariant under the mirror operation. Consequently,  $\sigma_h$ does not belong to the little group
of an arbitrary $\mathbf{k}$ or $\mathbf{q}$ point, and electronic states and phonon modes cannot, in general, be assigned a definite mirror parity throughout the entire Brillouin zone.

Mirror parity remains well defined only on the planes satisfying $\sigma_h\mathbf{k}=\mathbf{k}+\mathbf{G}$ (and analogously for $\mathbf{q}$), namely the mirror-invariant planes of the three-dimensional Brillouin zone. On these planes, Bloch states and phonon modes can again be chosen as eigenstates of $\sigma_h$, and the selection rule derived above applies directly whenever the initial state $\mathbf{k}$, the phonon wave vector $\mathbf{q}$, and the final state $\mathbf{k}+\mathbf{q}$ all belong to mirror-invariant subspaces.

Figure~\ref{fig:bands_FS_atoms_sigma_pi_MgB2} summarizes the electronic
structure of \ce{MgB2} and its classification in terms of orbital character and mirror parity. Panel~(a) shows the electronic band structure near the Fermi level, with the states colored
according to their projections onto the B $s$, $p_x$, $p_y$, and $p_z$ atomic orbitals, together with the corresponding projected and total density of states (DOS). The states crossing the Fermi level are predominantly of either $p_z$ character (green) or in-plane $p_x,p_y$ character (blue and
purple, respectively), corresponding to the conventional $\pi$- and $\sigma$-derived bands of \ce{MgB2}\cite{kong2001electron,PhysRevLett.86.4656,an2001superconductivity,choi2002first}.

Panel~(b) complements this orbital picture by showing the same electronic bands colored according to their exact eigenvalue under $\sigma_h$, wherever mirror parity is well defined. Along the portions of the high-symmetry path lying within the mirror-invariant planes
$k_z=0$ ($\Gamma$, M, K) and $k_z=\pi/c$ (A, L, H), each electronic state can be unambiguously classified as mirror even (red, labeled $\sigma$) or mirror odd (blue, labeled $\pi$). By contrast, along the
$\Gamma$--A segment, a generic wave vector $(0,0,k_z)$ is mapped by $\sigma_h$ onto $(0,0,-k_z)$ and is therefore not left invariant by the mirror operation. The electronic states along this segment
consequently have no well-defined mirror parity and are shown in gray, except at the endpoints $\Gamma$ and A, where mirror parity is again well defined.

It is worth noting that, wherever mirror parity is well defined, the
orbital-weight picture of panel~(a) shows a clear correspondence with
the exact symmetry classification of panel~(b). States dominated by  in-plane B $p_x$ and $p_y$ orbitals are predominantly mirror even,
whereas those with dominant B-$p_z$ character are mirror odd. Thus, on
the mirror-invariant planes, the conventional $\sigma$--$\pi$ orbital
classification of \ce{MgB2} closely matches the exact mirror-parity
classification.

The corresponding three-dimensional Fermi surface is shown in panel (c),
colored according to $\chi_\mathrm{orb}(\mathbf{k})$, the normalized
difference between the projected weights on atomic orbitals that are even and odd under $z\to-z$ about their own atomic center. Red and blue denote states dominated by even and odd orbitals, respectively, and intermediate colors denote mixed character (see SM for details). The quasi-two-dimensional
cylindrical sheets display a strongly predominant $\sigma$ character,
in agreement with the well-established orbital description of
\ce{MgB2}. In contrast, the more three-dimensional, interconnected
Fermi-surface sheets occupying the central region are predominantly
$\pi$-like. Their orbital character is not completely uniform,
however: while some regions exhibit an almost pure $\pi$ character
(dark blue), others appear in lighter shades of blue, indicating a
smaller imbalance between the $\pi$ and $\sigma$ orbital weights.

The Fermi surface colored according to the exact
$\sigma_h$ parity is shown in
Fig.~\ref{fig:bands_FS_atoms_sigma_pi_MgB2}(d). In contrast to the
orbital-character representation of panel~(c), mirror parity is defined
only along the intersections of the Fermi surface with the
mirror-invariant planes $k_z=0$ and $k_z=\pi/c$. The remaining
Fermi-surface states, shown in gray, lie at generic $k_z$ and therefore
cannot be assigned a definite mirror parity. Wherever the parity is
well defined, its classification closely follows the orbital-weight
picture of panel~(c): $\sigma$-dominated regions are mirror even,
whereas $\pi$-dominated regions are mirror odd.

The comparison between panels~(c) and (d) also highlights the
fundamental difference between a continuous orbital-character descriptor
and an exact symmetry eigenvalue. For example, along the central
three-dimensional Fermi-surface sheet intersecting the
$k_z=\pi/c$ (A--H--L) plane, $\chi_\mathrm{orb}(\mathbf{k})$ varies
substantially and indicates a weaker $\pi$ dominance in some regions.
Nevertheless, all of these states are exactly mirror odd wherever the
parity is defined. Thus, a reduction in the relative $\pi$ orbital
weight does not imply a corresponding weakening of the mirror
character: the former varies continuously, whereas the latter remains
a discrete symmetry eigenvalue.

We next examine how the electron--phonon matrix elements evolve as the
electronic wave vector moves away from a mirror-invariant plane.
Starting from $k_z=0$, where the electronic states possess exact mirror
parity and the selection rule applies rigorously, we increase $k_z$
while keeping the in-plane components $(k_x,k_y)$ fixed and track the
same electronic bands and their corresponding electron--phonon matrix
elements. At generic finite $k_z$, $\sigma_h$ maps
$(k_x,k_y,k_z)$ onto $(k_x,k_y,-k_z)$ rather than onto the same wave
vector. The electronic states therefore no longer possess a well-defined
mirror eigenvalue, and the matrix elements belonging to channels that
are symmetry forbidden at $k_z=0$ are no longer constrained to vanish.
This construction allows us to follow quantitatively how these
initially forbidden scattering channels become activated upon leaving
the mirror-invariant plane.

\begin{figure*}
\centering
\includegraphics[scale=0.265]{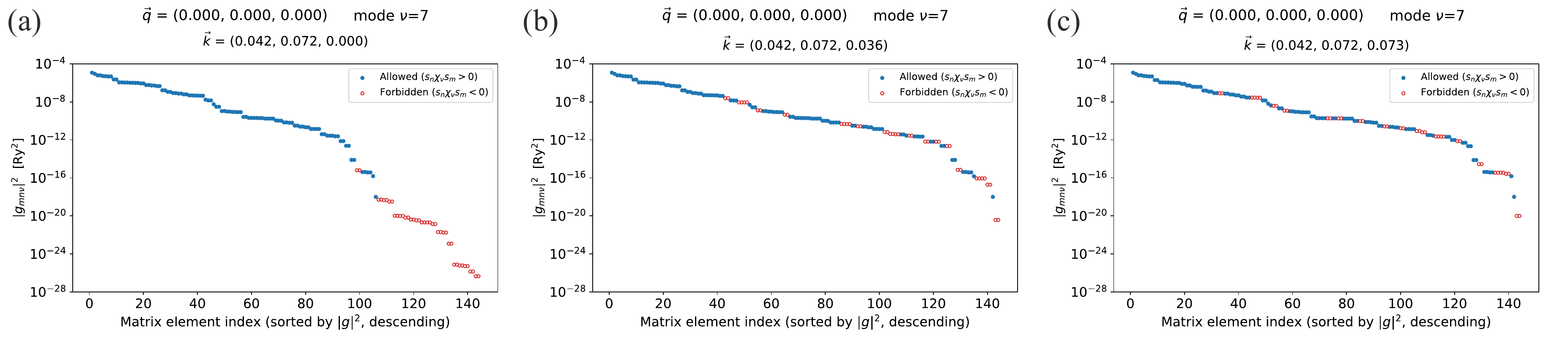}
\caption{Electron-phonon matrix elements $|g_{mn\nu}(\mathbf{k},\mathbf{q})|^2$
of \ce{MgB2} for the $E_{2g}$ mode ($\nu=7$) at $\mathbf{q}=\Gamma$, sorted in
descending order, for three $\mathbf{k}$ points with the same in-plane
components and increasing $k_z$:
(a) $k_z=0$, (b) $k_z=\pi/12c$, and (c) $k_z=\pi/6c$ (coordinates in units of
$2\pi/a$). All $12\times12$ band pairs are shown. Filled blue (open red)
symbols denote elements allowed (forbidden) by the sign of
$s_n(\mathbf{k})\,\chi_\nu(\mathbf{q})\,s_m(\mathbf{k}+\mathbf{q})$.
In (a), $\sigma_h$ belongs to the little group of $\mathbf{k}$ and $s_n$ is the
exact mirror eigenvalue: forbidden elements vanish to numerical precision
($\lesssim10^{-15}$~Ry$^2$). In (b) and (c), mirror parity is not defined and
$s_n$ is replaced by the projection-based estimator $\chi_\mathrm{proj}$
(see SM).}
\label{fig:diag_k1-3}
\end{figure*}

To assess how much the selection rule still constrains the EPC at a
generic point of the Brillouin zone, where it no longer applies
(e.g., $k_{z}\neq 0$, $\pi/c$), we consider a representative case: the \(E_{2g}\)
mode (\(\nu=7\)) at \(\mathbf{q}=\Gamma\), whose mirror parity remains
well defined, so that any departure from the rule originates solely
from the electronic states. Figure~\ref{fig:diag_k1-3} shows the
corresponding matrix elements for three \(\mathbf{k}\) points with
identical in-plane components and increasing \(k_z\). Panel~(a)
(\(k_z=0\)) is the baseline, where the rule holds exactly: the largest
forbidden elements lie more than ten orders of magnitude below the
largest allowed ones, and the ratio between their mean values is
\(\sim10^{-11}\), consistent with the numerical accuracy of DFPT.

In panels~(b) and (c), the electronic states no longer possess a
well-defined \(\sigma_h\) eigenvalue. To label them, we introduced the
projection-based estimator \(\chi_{\mathrm{proj}}\) (see SM), which
reduces to \(s_n(\mathbf{k})\) on the mirror-invariant planes and
assigns an effective mirror parity to each state elsewhere. This allows
us to follow the channels that evolve continuously from the forbidden
sector at \(k_z=0\). The outcome is twofold. On the one hand, the forbidden-like elements
grow rapidly once the rule is lifted, mixing with the allowed-like
ones, and the ratio between their mean values increases by about nine
orders of magnitude. On the other hand, the dominant elements remain
allowed-like, with the largest forbidden-like ones lying two to three
orders of magnitude below the largest allowed-like ones.

This behavior quantitatively demonstrates that the exact zeros imposed
by the selection rule are protected only when the participating
electronic states possess well-defined mirror eigenvalues. Once
$\mathbf{k}$ leaves the mirror-invariant plane, this symmetry
constraint no longer applies and the channels that were strictly
forbidden at $k_z=0$ can become active.

\section{Computational Details}
\label{sec:computational}
Geometry optimizations and electronic structure calculations were carried out using  density-functional theory (DFT)\cite{hohenbergkohn1964,PhysRev.140.A1133} as implemented in the Quantum Espresso (QE) package v7.5.1 \cite{giannozzi2009quantum,Giannozzi2017_QE,Lin2022_npjComputMater}. All these calculations were performed using the generalized gradient approximation (GGA) functional of Perdew-Burke-Ernzerhof revised for solids (PBEsol) \cite{PhysRevLett.100.136406}. The core electrons were simulated with the norm-conserving pseudopotentials from the Pseudo-Dojo database \cite{van2018pseudodojo,hamann2013optimized}. The number of valence electrons explicitly considered for each element is listed in Table S1 of the SI. To ensure the completeness of the basis, we set the plane-wave kinetic-energy cutoff to 100 Ry and the charge-density cutoff to 400 Ry. The reciprocal space was sampled using a $\Gamma$-centered Monkhorst-Pack mesh\cite{PhysRevB.40.3616} with a gaussian smearing width of 0.02 Ry. 

For the \ce{Sc2BCH4} monolayer, spurious interactions with periodic images  along the aperiodic direction were avoided by imposing 15~\AA~of vacuum  together with a cutoff of the Coulomb interaction in that direction  \cite{PhysRevB.96.075448}. To mitigate the well known delocalization errors of GGA\cite{doi:10.1021/ct800531s}, GGA+$U$ calculations were performed within the Dudarev's formulation\cite{PhysRevB.57.1505} using atomic projectors. Hubbard $U$ parameters were obtained self-consistently via the density-functional perturbation theory (DFPT) framework\cite{PhysRevB.98.085127,TIMROV2022108455}. The resulting elemental $U$ values are reported in Tables S2 and S3 in the SI. In contrast, bulk \ce{MgB2} does not require a Hubbard 
$U$ correction and was treated at the plain GGA level, without any vacuum 
spacing.

We employed our modified in-house Quantum ESPRESSO package to perform DFPT phonon and electron--phonon coupling calculations, consistently including the GGA+$U$ corrections for \ce{Sc2BCH4}. The dynamical matrices and phonon-induced perturbations were computed on $6\times6\times1$
(\ce{Sc2BCH4}) and $6\times6\times6$ (\ce{MgB2}) $\mathbf{q}$-point grids,
and the electron--phonon matrix elements were evaluated on
$\Gamma$-centered $48\times48\times1$ and $24\times24\times24$ $\mathbf{k}$-point grids, respectively. For the symmetry analysis, the matrix elements $g_{mn\nu}(\mathbf{k},\mathbf{q})$ were computed directly
from DFPT without interpolation, so that their values, including the
symmetry-forbidden ones, are not altered by any interpolation, and were written out individually. 

The critical temperature $T_c$ of \ce{Sc2BCH4} was estimated by solving the anisotropic 
Migdal--Eliashberg equations~\cite{migdal1958interaction,eliashberg1960interactions} 
using the EPW code v6.1~\cite{giustino2007electron,lee2023electron,ponce2016epw,noffsinger2010epw,margine2013anisotropic,yang2025first}. The electron--phonon matrix 
elements were first computed on coarse electron and phonon grids of 
$24\times24\times1$ and $8\times8\times1$ $\mathbf{k}$- and $\mathbf{q}$-points, 
respectively, and subsequently interpolated onto fine grids of 
$80\times80\times1$ and $40\times40\times1$ $\mathbf{k}$- and $\mathbf{q}$-points 
within EPW. Convergence tests and validation of the Wannier interpolation are 
reported in Fig.~S4 of the SI.  We employed a Matsubara frequency cutoff of 
$w_{\mathrm{scut}}=1.5$~eV. The electron--electron 
interaction was described through the Morel--Anderson pseudopotential, with the 
effective Coulomb parameter set to $\mu^{*}=0.10$, as commonly assumed for 
hydrides. The Eliashberg spectral function $\alpha^2F(\omega)$ was computed 
with a Fermi window of $f_{\mathrm{thick}}=0.5$~eV and a smearing of 
$\delta_{\mathrm{q}}=0.5$~meV for the sum over $\mathbf{q}$.

\section{Conclusions}
We have derived an exact mirror-parity selection rule for
electron--phonon scattering in two-dimensional crystals having a
horizontal mirror plane, $\sigma_h$. The matrix element
$g_{mn,\nu}(\mathbf{k},\mathbf{q})$ vanishes unless
$s_n(\mathbf{k})\chi_\nu(\mathbf{q})
s_m(\mathbf{k}+\mathbf{q})=+1$. Thus, mirror-even phonons couple
electronic states of the same mirror parity, whereas mirror-odd phonons
couple states of opposite parity. Because $\sigma_h$ leaves every
in-plane wave vector invariant, this constraint applies throughout the
Brillouin zone of a strictly two-dimensional crystal.

First-principles calculations for \ce{Sc2BCH4} confirm that
symmetry-forbidden matrix elements vanish up to numerical precision,
while symmetry-allowed channels may span a wide range of coupling
strengths. Anisotropic Migdal--Eliashberg calculations further yield
$T_c\approx100$~K. 

In bulk \ce{MgB2}, the same selection
rule remains exact only on mirror-invariant planes of the
three-dimensional Brillouin zone. Away from these planes, the
electronic mirror eigenvalues cease to be defined and scattering
channels that are symmetry forbidden on the invariant planes become
active. These results establish horizontal mirror symmetry as an exact
filter of electron--phonon scattering channels and provide a general
framework for identifying symmetry-enforced zeros in electron--phonon
matrix elements.
\begin{acknowledgments}
This research was funded by the Association Nationale de la Recherche 
under grant ANR-22-CE50-0014 and by ECOS-Sud under grants C1709 and 
C21E06/ECOS210019.  This work was funded by ANID thorough the Center for Nanoscience and 
Nanotechnology CEDENNA CIA250002  and Fondecyt grants 1231487, 1260378. N.F.B. thanks the ED388 
doctoral school for his doctoral grant.  C.C acknowledges Universidad Nacional de Colombia, Sede Medellín.  This work was granted access to 
the HPC resources of the SACADO MeSU platform at Sorbonne Université, 
and computational time through GENCI projects A0160915069, A0160815101, 
and A0190915069 is greatly acknowledged. Powered@NLHPC: this research 
was partially supported by the supercomputing infrastructure of the 
NLHPC (CCSS210001).
\end{acknowledgments}

\bibliography{apssamp}

\end{document}